\documentclass[11pt]{article}
\usepackage[utf8]{inputenc}
\usepackage{amsmath,amsthm,amsfonts,amssymb,amscd}
\usepackage{multirow,booktabs}
\usepackage[table]{xcolor}
\usepackage[font=small,labelfont=bf]{caption}
\usepackage{fullpage}
\usepackage{lastpage}
\usepackage{enumitem}
\usepackage{fancyhdr}
\usepackage{mathrsfs}
\usepackage{wrapfig}
\usepackage{setspace}
\usepackage{calc}
\usepackage{multicol}
\newsavebox{\gridbox}
\usepackage{cancel}
\usepackage{cases}
\numberwithin{equation}{section}
\usepackage[retainorgcmds]{IEEEtrantools}
\usepackage[margin=3cm]{geometry}

\usepackage{empheq}
\usepackage{framed}
\usepackage[most]{tcolorbox}
\usepackage{cite}
\usepackage[colorlinks=true,linkcolor=blue,citecolor=blue,urlcolor=blue]{hyperref}
\colorlet{shadecolor}{orange!15}
\theoremstyle{definition}

\begin{document}

\title{\Large{\bf 
Effects of Self-Interactions on the Bifurcations in Boson Stars
} }

\author{
{\large Chen Liang}
\vspace*{0.2cm}
\\
{\small Departamento de Matem\'atica da Universidade de Aveiro and } 
\\ {\small  Centre for Research and Development  in Mathematics and Applications (CIDMA),} 
\\ {\small    Campus de Santiago, 3810-193 Aveiro, Portugal}
}
\maketitle 

\begin{abstract}
Radially excited mini-boson stars have zero modes under static axisymmetric perturbations, at which they bifurcate into axisymmetric configurations with chain-like and ring-like matter distributions. We study how self-interactions affect the $\ell=2$ zero modes of spherical boson stars with one radial node, and construct the chains and rings that bifurcate from them. We consider massive and solitonic boson stars, which have a repulsive quartic self-interaction and a Q-ball type potential, respectively. For massive boson stars, the zero mode persists for all values of the quartic coupling we considered. As the coupling grows, it moves onto the candidate stable branch, so that this branch cannot be entirely stable against $\ell=2$ perturbations, even though sufficiently strong self-interactions are known to quench the radial instability of these stars. For solitonic boson stars, the bifurcations depend strongly on the gravitational coupling. As it decreases, a single family of spherical stars can have multiple zero modes, at which chains and rings alternate, and below a critical coupling the zero modes disappear. In flat spacetime, the rings still exist as static ring-like Q-balls, whereas no chains are found, which suggests that chains need gravity to exist.
\end{abstract}

\tableofcontents

 \section{Introduction}
Q-balls are non-topological solitons formed by a complex scalar field with a suitable self-interaction in flat spacetime~\cite{Rosen:1968mfz,Friedberg:1976me,Coleman:1985ki,Lee:1991ax}. When the scalar field is coupled to gravity, localized solutions exist even without self-interactions~\cite{Kaup:1968zz,Ruffini:1969qy}. These are the simplest boson stars (BSs), usually called \emph{mini-BSs}~\cite{Jetzer:1991jr,Schunck:2003kk,Liebling:2012fv,Shnir:2022lba}. Among self-interacting BSs, two commonly studied types differ in whether gravity is needed to bind them. \emph{Massive BSs}, with a repulsive quartic self-interaction, can reach much larger masses than mini-BSs~\cite{Colpi:1986ye}, but they are still bound only by gravity, since this self-interaction does not support solitons in flat spacetime. In contrast, \emph{solitonic BSs}, which have a Q-ball type potential with an attractive quartic and a repulsive sextic term\footnote{The sextic potential of solitonic BSs is the simplest polynomial potential that supports Q-balls~\cite{Volkov:2002aj}. In principle, higher-order terms could also be included~\cite{Heeck:2022iky}, but we do not consider them here.}, can be regarded as gravitating Q-balls~\cite{Friedberg:1986tq}. They reduce to Q-balls in the limit of vanishing gravitational coupling and to mini-BSs in the limit of strong coupling~\cite{Kleihaus:2005me,Kleihaus:2007vk}.

BSs have been widely studied in the context of scalar field dark matter~\cite{Hu:2000ke,Suarez:2013iw,Hui:2016ltb} and as black hole mimickers~\cite{Cardoso:2019rvt,Olivares:2018abq,Herdeiro:2021lwl}. However, unlike static vacuum black holes, which must be spherically symmetric if they are asymptotically flat and regular on and outside the event horizon~\cite{Israel:1967wq}, static BSs are not necessarily spherically symmetric. The simplest non-spherical example is the dipolar BS~\cite{Yoshida:1997nd,Cunha:2022tvk}, which can be regarded as two BSs with opposite phases, kept in equilibrium by the balance between their scalar repulsion and gravitational attraction~\cite{Palenzuela:2006wp}. Such multi-component configurations also include chains with more constituents along the symmetry axis~\cite{Herdeiro:2021mol,Sun:2022duv} and ring-like BSs~\cite{Liang:2025myf}. Moreover, in analogy with the orbitals of the hydrogen atom, static BSs can exhibit multipolar structures, and the generic solutions possess only discrete symmetries~\cite{Herdeiro:2020kvf}. Similar multi-component structures also exist for rotating BSs~\cite{Kleihaus:2007vk,Gervalle:2022fze,Sun:2023ord,Sun:2025eis,Liang:2026vls} and for solitons in other models, both with and without gravity~\cite{Shnir:2021gki,Jaramillo:2024cus,Loiko:2020htk}.

Similar to the hydrogen atom, spherical BSs (SBSs) also have excited states, in which the scalar field has radial nodes~\cite{Friedberg:1986tp}. Excited SBSs have been proposed as models of galactic dark matter halos, since they give flatter rotation curves than the fundamental SBSs~\cite{Lee:1995af,Brito:2023fwr}, so their stability is not only of theoretical interest. Without self-interactions, the excited SBSs are unstable~\cite{Lee:1988av,Jetzer:1989vs,Balakrishna:1997ej}, even in the part of the family of solutions from the Newtonian limit up to the maximum mass. This part is stable for the fundamental SBSs~\cite{Gleiser:1988rq,Gleiser:1988ih,Lee:1988av,Seidel:1990jh}, and for the excited SBSs it is referred to as the \emph{candidate stable branch}~\cite{Sanchis-Gual:2021phr}. With a sufficiently strong quartic self-interaction, the excited SBSs can be stabilized in spherically symmetric evolutions~\cite{Sanchis-Gual:2021phr,Brito:2023fwr} (see also Ref.~\cite{Hao:2026ogd} for a linear analysis of their radial modes). A similar stabilizing effect of self-interactions was found for spinning~\cite{DiGiovanni:2020ror,Siemonsen:2020hcg} and dipolar BSs~\cite{Ildefonso:2023qty}. Beyond spherical symmetry, however, both a linear analysis in the Newtonian limit~\cite{Nambo:2024gvs} and fully nonlinear 3+1 evolutions~\cite{Brito:2025rld} show that the excited SBSs, like excited Q-balls in flat spacetime~\cite{Chen:2025oxo}, are generically unstable against non-spherical perturbations, and in the evolutions the instability appears earlier for stronger self-interactions.

Non-spherical perturbations also connect the excited SBSs with the static axisymmetric BSs (ABSs) discussed above. In our previous work~\cite{Liang:2025myf}, we found that the excited SBSs have zero modes under static axisymmetric perturbations, at which they bifurcate into ABSs with chain-like and ring-like matter distributions, which we call \emph{chains} and \emph{rings}. This explains why chains with an odd number of constituents merge with excited SBSs, as observed in Refs.~\cite{Herdeiro:2021mol,Sun:2022duv}. Structures resembling chains and rings have also been seen in collisions of excited SBSs~\cite{Brito:2025rld,Ge:2026wzh}. Moreover, at a zero mode the frequency of a non-spherical perturbation mode vanishes, and the SBSs on its two sides generically differ by one unstable mode. Therefore, the zero modes also mark a change in the stability of SBSs against non-spherical perturbations.

The zero modes in Ref.~\cite{Liang:2025myf} were obtained for mini-BSs. Since a strong enough quartic self-interaction stabilizes the excited SBSs in spherically symmetric evolutions~\cite{Sanchis-Gual:2021phr,Brito:2023fwr}, we speculated there that it might also remove the zero modes and thus suppress the bifurcations. In this work, we test this idea for massive BSs. We also consider solitonic BSs to see how the bifurcations depend on the strength of gravity relative to the self-interactions. In both models, we consider the SBSs with one radial node, which are the first excited states. We follow their $\ell=2$ zero modes as the couplings vary, and construct the chains and rings that bifurcate from them. We find no sign that the quartic self-interaction removes the zero mode. Instead, for strong self-interactions, the zero mode moves onto the candidate stable branch, so that this branch cannot be entirely stable against $\ell=2$ perturbations. For solitonic BSs, depending on the gravitational coupling, a single SBS family can have two or more zero modes, and a chain branch and a ring branch bifurcate from each of them. Finally, we find that the rings still exist in flat spacetime, as static ring-like Q-balls, which, to our knowledge, have not been reported before.

The paper is organized as follows. In Sec.~\ref{sec:framework}, we introduce the two models and the ansatz for the SBSs and ABSs, and give the field equations and the equations for static axisymmetric perturbations of the SBSs. Section~\ref{sec:bc} presents the boundary conditions and the physical quantities used to describe the solutions. The scaling symmetries and the numerical methods are described in Sec.~\ref{sec:numerics}. In Sec.~\ref{sec:results}, we present the numerical results, first for massive BSs and then for solitonic BSs. Our conclusions are given in Sec.~\ref{sec:conclusions}.

\section{Framework}\label{sec:framework}
\subsection{The models and ansatz}
We consider a complex scalar field minimally coupled to Einstein's gravity. The action of this model is given by
\begin{equation}\label{eq1.1}
    S = \int d^4x \sqrt{-g}\left( \frac{R}{16 \pi G} - g^{ab}\Psi^*_{,a}\Psi_{,b} - U(|\Psi|^2) \right)\,,
\end{equation}
where $U(|\Psi|^2)$ is the potential of the scalar field. Varying the action with respect to the metric and the scalar field leads to the following equations:
\begin{subequations}\label{equek}
\begin{align}
\label{eque}
&
E_{ab} =R_{ab} - \frac{1}{2}g_{ab}R-2\alpha^2 T_{ab}=0\,, \\
\label{equk}
&
\left(\nabla^2 - \frac{\partial U}{\partial|\Psi|^2} \right)\Psi = 0\,, 
\end{align}
\end{subequations}
where $\alpha^2=4\pi G$. The energy-momentum tensor reads
\begin{eqnarray} \label{energy-momentum}
T_{ab} = \Psi^*_{,a}\Psi_{,b} + \Psi^*_{,b}\Psi_{,a} - g_{ab}\left[ \frac{1}{2}g^{cd} \left( \Psi^*_{,c}\Psi_{,d} + \Psi^*_{,d}\Psi_{,c} \right) + U(|\Psi|^2) \right]\, .
\end{eqnarray} 
To study the bifurcations of SBSs into ABSs, we consider a general ansatz for the line element:
\begin{eqnarray}
ds^2= g_{tt}(r,\theta)dt^2 + g_{rr}(r,\theta)dr^2 + g_{\theta\theta}(r,\theta)d\theta^2 + g_{\varphi\varphi}(r,\theta)d\varphi^2\,,
\end{eqnarray}
where the metric functions for SBSs are 
\begin{eqnarray}\label{eq1.5}
g_{tt}=-N(r)\sigma^2(r)\,,\quad g_{rr}=\frac{1}{N(r)}\,,\quad g_{\theta\theta}=r^2\,,\quad g_{\varphi\varphi}=r^2\sin^2{\theta}\,,
\end{eqnarray}
with $N(r)=1-2m(r)/r$. For the ABSs, the metric functions are given by
\begin{eqnarray}\label{eq1.6}
    g_{tt}=-e^{2F_0(r,\theta)}\,,\quad g_{rr}=e^{2F_1(r,\theta)}\,,\quad g_{\theta\theta}=r^2e^{2F_1(r,\theta)}\,,\quad g_{\varphi\varphi}=r^2\sin^2{\theta}e^{2F_2(r,\theta)}\,.
\end{eqnarray}
Finally, the matter field has the following form:
\begin{eqnarray}\label{eq1.7}
    \Psi=\psi(r,\theta) e^{-i\omega t}\,,
\end{eqnarray}
where the function $\psi$ depends only on $r$ for the SBSs.

We consider two types of self-interactions:
    \begin{subnumcases}{U=}
        \label{quartic}
        ~\mu^2|\Psi|^2 + \lambda|\Psi|^4 & \text{(massive)}\,,\\
        \notag
        \\
        \label{q-ball}
        ~\mu^2|\Psi|^2 - \lambda|\Psi|^4 + \nu|\Psi|^6 & \text{(solitonic)}\,,
    \end{subnumcases}
where $\mu$ is the mass of the scalar field, and $\lambda$ and $\nu$ are non-negative coupling constants. 

The theory \eqref{eq1.1} possesses a global $U(1)$ symmetry under the transformation $\Psi \to e^{i\chi}\Psi$, with $\chi$ a constant. This global invariance implies the existence of a 4-current
\begin{align}
j^{a}
  = -i\left( \Psi^* \partial^{a}\Psi - \Psi \partial^{a}\Psi^*  \right),
  \label{eq:current}
\end{align}
which is conserved as a consequence of the field equation \eqref{equk}. Integrating the time component $j^t$ over a spacelike hypersurface $\Sigma$ yields the conserved Noether charge
\begin{align}
 Q = \int_{\Sigma} d^3 x \sqrt{-g}\, j^{t} \,.
  \label{eq:Q}
\end{align}

\subsection{The equations of motion}\label{sec2.2}
For SBSs, inserting the ansatz (\ref{eq1.5}) and (\ref{eq1.7}) into (\ref{equek}) reduces the field equations to the following set of ordinary differential equations for $m(r)$, $\sigma(r)$ and $\psi(r)$:
\begin{subequations}\label{bgequ_spherical}
  \begin{align}
    & \psi''+ 
    \left(\frac{N'}{N}+\frac{\sigma'}{\sigma}+\frac{2}{r}\right)\psi'+
    \left(\frac{\omega^2}{N^2 \sigma^2}-\frac{1}{N}\frac{\partial U}{\partial|\Psi|^2}\right)\psi=0\,,
    \label{bgequ_spherical_1}\\
    & m'-
    \alpha^2 r^2\left(  N \psi'^2+
     \frac{\omega^2}{N \sigma^2}\psi^2 + U \right)=0\,,
     \label{bgequ_spherical_2}\\
    & \sigma'- 
    2\alpha^2 r\left[  \sigma \psi'^2+
    \frac{ \omega^2 \psi^2}{N^2 \sigma} \right] =0\,.
    \label{bgequ_spherical_3}
\end{align}
\end{subequations}
For ABSs, substituting the ansatz (\ref{eq1.6}) and (\ref{eq1.7}) into (\ref{equek}) and performing appropriate simplifications, we obtain the following set of equations for constructing ABS solutions:
\begin{subequations}\label{bgequ_axial}
  \begin{align}
& 
\begin{split}
e^{2 F_1}\psi \left( \omega^2 e^{-2 F_0} - \frac{\partial U}{\partial |\Psi|^2}\right)+
\frac{\psi _{,\theta} \left(F_{0,\theta}+F_{2,\theta}+\cot\theta\right)}{r^2}+\psi_{,r} \left(F_{0,r}+F_{2,r}+\frac{2}{r}\right)+\\
\frac{\psi _{,\theta \theta}}{r^2}+\psi _{,rr}=0\,,
\end{split}
\label{bgequ_axial_1}\\[6pt]
& \begin{split}
    2\alpha^2 e^{2 F_1}   \left(U  -2\omega^2 e^{-2 F_0} \psi ^2 \right)+\frac{F_{0,\theta}\left(F_{0,\theta}+F_{2,\theta}+\cot\theta\right)}{r^2}+F_{0,r}\left(F_{0,r}+F_{2,r}+\frac{2}{r}\right)+\\
    \frac{F_{0,\theta \theta}}{r^2}+F_{0,rr}=0\,,
\end{split}
\label{bgequ_axial_2}\\[6pt]
& \begin{split}
    2\alpha^2 \left(\psi ^2 \omega^2 e^{2 F_1-2 F_0}+\frac{\psi _{,\theta}^2}{r^2}+\psi_{,r}^2\right)-\frac{F_{0,\theta} \left(F_{2,\theta}+\cot\theta\right)}{r^2}-F_{0,r} \left(F_{2,r}+\frac{1}{r}\right)+\frac{F_{1,\theta \theta}}{r^2}+\\
    \frac{F_{1,r}}{r}+F_{1,rr}=0\,,
\end{split}
\label{bgequ_axial_3}\\[6pt]
& \begin{split}
   \frac{F_{2,\theta} \left(F_{0,\theta}+2 \cot\theta+ F_{2,\theta} \right)}{r^2}+F_{2,r}\left(F_{0,r}+F_{2,r}+\frac{3}{r}\right)+\frac{F_{0,\theta} \cot\theta}{r^2}+\frac{F_{0,r}}{r}+ 2\alpha^2  e^{2 F_1} U+\\
\frac{F_{2,\theta\theta }}{r^2}+F_{2,rr}=0\,.
\end{split}
\label{bgequ_axial_4}
  \end{align}
\end{subequations}
In addition, two further Einstein equations, $E_r^r - E_\theta^\theta=0$ and $E_r^\theta=0$, serve as constraint equations and are used to verify the accuracy of the ABS solutions.

\subsection{Axisymmetric perturbations}

In our previous work~\cite{Liang:2025myf}, we found that excited SBSs can bifurcate into ABSs under the following perturbations:
\begin{eqnarray}\label{perturbation}
&& \qquad\qquad\qquad\qquad\quad
\bar{\psi}(r,\theta)  = \psi(r) + \epsilon \psi_1(r)P_\ell(\cos \theta), 
\nonumber \\
&&
\bar{g}_{tt}  =- N(r) \sigma^2(r)(1 - \epsilon H(r)P_\ell(\cos \theta)),\quad \bar{g}_{rr}  = \frac{1}{N(r)}(1 + \epsilon H(r)P_\ell(\cos \theta)),
\\
&&
\bar{g}_{\theta \theta} = r^2 (1 + \epsilon K(r)P_\ell(\cos \theta)), \quad
\bar{g}_{\varphi \varphi} = r^2 \sin^2 \theta (1 + \epsilon K(r)P_\ell(\cos \theta)), 
\nonumber
\end{eqnarray}
where $\epsilon$ is a small parameter, $P_\ell(\cos \theta)$ is the Legendre polynomial of degree $\ell$, and $H(r)$, $K(r)$ and $\psi_1(r)$ are the perturbation functions in the Regge-Wheeler gauge~\cite{Regge:1957td}. Since we focus only on the zero modes of the SBSs, which indicate the bifurcations, time-dependent perturbations are not considered in this work. 

Substituting (\ref{perturbation}) into (\ref{eque}) and expanding in $\epsilon$, we obtain
\begin{equation}\label{equeper}
  E_{ab} + \epsilon E_{ab}^{(1)} + \epsilon^2 E_{ab}^{(2)} + \cdots = 0 \,.
\end{equation}
The zeroth-order equations are satisfied by the background solution, so we keep only the first-order terms $E_{ab}^{(1)}$. We combine the $(r,r)$, $(r,\theta)$ and $(\theta,\theta)$ components to eliminate $K$ and use the background equations. The first-order Einstein and Klein-Gordon equations then reduce to two coupled equations for $H$ and $\psi_1$:
\begin{subequations}\label{p-eq-system}
  \begin{align}
    \mathcal{D} H + V_{11} H + V_{12}\, \psi_1 &= 0 \,, \label{p-eq-metric} \\
    \mathcal{D} \psi_1 + V_{21} H + V_{22}\, \psi_1 &= 0 \,, \label{p-eq-scalar}
  \end{align}
\end{subequations}
where both equations share the radial operator
\begin{equation}\label{radial-operator}
  \mathcal{D} \equiv \frac{1}{r^2 N \sigma}\,\frac{d}{dr}\left( r^2 N \sigma\,\frac{d}{dr} \right) .
\end{equation}
The coefficients depend only on the background solution and read
\begin{subequations}\label{V-coefficients}
  \begin{align}
    V_{11} &= \frac{2-\ell(\ell+1)}{r^2 N} - \frac{N + 8\alpha^2 r^2 U}{r^2 N} - \left( \frac{1 - 2\alpha^2 r^2 U}{rN} \right)^2 \notag \\
    &\quad - \frac{4\alpha^2 \omega^2 \psi^2}{N^2 \sigma^2} \left( \frac{1 - 2\alpha^2 r^2 U}{N} - 5 + 2\alpha^2 r^2 \psi'^2 + \frac{\alpha^2 \omega^2 r^2 \psi^2}{N^2 \sigma^2} \right) \notag \\
    &\quad + 4\alpha^2 \psi'^2 \left( 1 - \frac{1 - 2\alpha^2 r^2 U}{N} \right) - 4\alpha^4 r^2 \psi'^4 \,, \\
    V_{22} &= -\frac{\ell(\ell+1)}{r^2 N} + \frac{\omega^2}{N^2 \sigma^2} - 8\alpha^2 \psi'^2 - \frac{1}{N} \left( \frac{\partial U}{\partial |\Psi|^2} + 2\psi^2\, \frac{\partial^2 U}{\partial (|\Psi|^2)^2} \right) , \\
    V_{21} &= \left( \frac{2\omega^2}{N^2 \sigma^2} - \frac{1}{N} \frac{\partial U}{\partial |\Psi|^2} \right) \psi - \left( \frac{1}{r} - \frac{1}{rN} - \frac{2\alpha^2 \omega^2 r}{N^2 \sigma^2}\, \psi^2 + \frac{2\alpha^2 r}{N}\, U \right) \psi' + 2\alpha^2 r\, \psi'^3 \,, \\
    V_{12} &= 8\alpha^2 V_{21} \,, \label{V-relation}
  \end{align}
\end{subequations}
where a prime denotes $d/dr$, and $U$ and its derivatives are evaluated on the background, $|\Psi|^2 = \psi^2$. Equation~(\ref{V-relation}) shows that $\psi_1$ sources $H$ only at order $\alpha^2$, so the metric perturbation decouples in the flat-spacetime limit $\alpha \to 0$.

SBSs with nonzero node number can only bifurcate into ABSs with the same parity as the SBSs, so only even values of $\ell$ are relevant here. For simplicity, we consider only the SBSs with one node and the quadrupole perturbation $\ell=2$. From now on, all SBSs have one node unless stated otherwise. 

\section{The boundary conditions and physical quantities}\label{sec:bc}
\subsection{Boundary conditions}
To construct the SBSs and ABSs numerically, appropriate boundary conditions must be imposed on the equations of Sec.~\ref{sec:framework}.

Regularity of the SBSs and ABSs requires the following conditions at $r=0$:
\begin{eqnarray}
  \partial_r \psi|_{r=0} = 0\,,\quad m(0) =0\,,\quad \sigma(0)= \sigma_0\,,\quad \partial_r F_{0,1,2}|_{r=0} = 0\,.
  \end{eqnarray}
Asymptotic flatness requires the following boundary conditions at infinity:
\begin{eqnarray}
  \psi|_{r=\infty} = 0 \,,\quad m(\infty) =M\,,\quad \sigma(\infty)= 1\,,\quad F_{0,1,2}|_{r=\infty} = 0\,.
  \end{eqnarray} 
In addition, for the ABSs, regularity on the symmetry axis requires 
\begin{eqnarray}
    \partial_\theta \psi|_{\theta=0,\pi} = 0 \,,\quad\partial_\theta  F_{0,1,2}|_{\theta=0,\pi} = 0\,,\quad F_1|_{\theta=0,\pi}=F_2|_{\theta=0,\pi}\,.  
\end{eqnarray} 
Since we consider only $\ell=2$ perturbations, the ABSs that bifurcate from the SBSs have the same parity as the SBSs. Thus the following boundary conditions are imposed at $\theta=\pi/2$ due to the $\mathbb{Z}_2$ symmetry of the ABSs:
\begin{eqnarray}
    \partial_\theta \psi|_{\theta=\frac{\pi}{2}}= 0 \,,\quad \partial_\theta F_{0,1,2}|_{\theta=\frac{\pi}{2}} = 0\,.
\end{eqnarray} 
For the perturbation equations (\ref{p-eq-system}), the background approaches Minkowski spacetime at infinity, so the perturbation functions satisfy 
\begin{eqnarray}
    H(\infty)= 0\,, \qquad 
    \psi_1(\infty)= 0\,.
\end{eqnarray}
Near the origin, regular solutions of (\ref{p-eq-system}) behave as
\begin{equation}\label{p-series-scalar}
  H(r) = h_\ell\, r^\ell + \mathcal{O}(r^{\ell+2}) \,, \qquad
  \psi_1(r) = p_\ell\, r^\ell + \mathcal{O}(r^{\ell+2}) \,,
\end{equation}
where $h_\ell$ and $p_\ell$ are constants. Since the perturbation equations are linear and homogeneous, one of these constants can be set to unity without loss of generality. In our numerical scheme, however, the amplitude is fixed instead by the normalization \eqref{H-normalization} imposed at an intermediate radius, which also supplies the extra condition that turns the search for zero modes into an eigenvalue problem for $\omega$. Both $h_\ell$ and $p_\ell$ are then determined by the solution.
\subsection{Physical quantities}
The asymptotic behavior as $r\rightarrow\infty$ of the metric functions $N(r)$ and $F_0(r,\theta)$ is 
\begin{eqnarray}\label{eq2.7}
   N(r)=1-\frac{2MG}{r}+\cdots \,, \qquad \qquad  F_0(r,\theta)=-\frac{MG}{r}+\cdots\,,
\end{eqnarray}
where $M$ is the ADM mass, which can also be computed as a Komar integral,
\begin{equation}\label{eq:Komar}
  M = \int \left( T_a^{a} - 2T_t^{t} \right) \sqrt{-g}\; dr\, d\theta\, d\varphi \,.
\end{equation}
For our ansatz the integrand simplifies to $T_a^{a} - 2T_t^{t} = 2\left( 2\omega^2\psi^2/(-g_{tt}) - U \right)$, so that
\begin{equation}\label{eq:M}
  \begin{aligned}
    \text{SBS:} \quad & M = 8\pi \int_0^\infty dr\, \frac{r^2}{N\sigma} \left( 2\omega^2\psi^2 - N\sigma^2 U \right) , \\
    \text{ABS:} \quad & M = 4\pi \int_0^\infty dr \int_0^\pi d\theta\; r^2 \sin\theta\, e^{2F_1+F_2-F_0} \left( 2\omega^2\psi^2 - e^{2F_0} U \right) .
  \end{aligned}
\end{equation}
Likewise, with $j^t = 2\omega\psi^2/(-g_{tt})$, the Noether charge \eqref{eq:Q} reads
\begin{equation}\label{eq:Qi}
  \begin{aligned}
    \text{SBS:} \quad & Q = 8\pi \int_0^\infty dr\, \frac{r^2}{N\sigma}\, \omega\psi^2 , \\
    \text{ABS:} \quad & Q = 4\pi \int_0^\infty dr \int_0^\pi d\theta\; r^2 \sin\theta\, e^{2F_1+F_2-F_0}\, \omega\psi^2 .
  \end{aligned}
\end{equation}
The higher-order terms in the asymptotic expansion of the metric functions encode the deviation from spherical symmetry:
\begin{eqnarray}
    \label{quadrupole1}
  &&  F_0(r,\theta)= -\frac{M G}{r} + \frac{1}{r^3}\left( b_1 - \frac{12b_1 - 4a_5 M +M^3}{4} \cos^2 \theta \right) + \cdots \,,\quad  \\
 &&   F_2(r,\theta) = \frac{M G}{r} + \frac{4a_5-M^2}{4r^2} +\cdots\,,
    \end{eqnarray} 
where $b_1$ and $a_5$ are constants determined by the solution. The \textit{quadrupole moment} is encoded in this expansion and is given by~\cite{Herdeiro:2015gia}
\begin{eqnarray}
  \label{quadrupole3}
  M_2 = -\nu_2 - \frac{4}{3}\left( \frac{1}{4} + \frac{B_0}{M^2} \right)M^3  \,,
  \end{eqnarray} 
with 
\begin{eqnarray}
  \label{quadrupole4}
  B_0 = a_5 - \frac{M^2}{4}\,,\quad \nu_2 = \frac{-12b_1 + 4 a_5 M - M^3}{6}\,.
  \end{eqnarray}
The quadrupole moment vanishes for SBSs. Since it measures how much the matter distribution deviates from spherical symmetry, it clearly shows the bifurcations in the solution space. We also analyze the matter distribution through the energy density
\begin{eqnarray}
  \label{energy-density}
  \rho = - T_t^t =
  U + e^{-2F_0}\omega^2\psi^2 +
\frac{e^{-2F_1}\left( \psi_{,\theta}^2 + r^2 \psi_{,r}^2 \right)}{r^2}
\,.
  \end{eqnarray}

\section{Scaling symmetries and numerical method}\label{sec:numerics}
The model (\ref{eq1.1}) possesses scaling symmetries, which we use to rescale the field equations in Sec.~\ref{sec2.2} for numerical convenience. For massive BSs we introduce the dimensionless quantities
\begin{eqnarray}
    \label{scalingq}
 \tilde{\psi}=\sqrt{4\pi G}\psi
 \,, \qquad 
 \tilde{r}=\mu r
 \,, \qquad 
 \tilde{\omega}=\frac{\omega}{\mu}
 \,, \qquad 
 \tilde{\lambda}=\frac{\lambda}{4\pi G \mu^2}
 \,.
\end{eqnarray} 
This rescaling amounts to setting $\alpha=\mu=1$ and $\lambda\to\tilde{\lambda}$ in the field equations. We analyze the impact of the quartic self-interaction on massive BSs by scanning the coupling $\tilde{\lambda}$.

For solitonic BSs we introduce instead
\begin{eqnarray}
    \label{scalingQ}
 \tilde{\psi}=\frac{\nu^{1/4}\sqrt{\mu_0}}{\sqrt{\mu}}\psi
 \,, \qquad 
 \tilde{r}=\frac{\mu}{\mu_0} r
 \,, \qquad 
 \tilde{\omega}=\frac{\mu_0}{\mu}\omega
 \,, \qquad \tilde{\lambda}=\frac{\mu_0}{\sqrt{\nu}\mu}\lambda
 \,, \qquad \tilde{\alpha}=\frac{\sqrt{\mu}}{\nu^{1/4}\sqrt{\mu_0}}\alpha\,,
\end{eqnarray}
where $\mu_0>0$ is an arbitrary constant. In these variables the field equations keep their form, with $\mu$, $\lambda$ and $\alpha$ replaced by $\mu_0$, $\tilde{\lambda}$ and $\tilde{\alpha}$, and with $\nu=1$. In particular, the potential (\ref{q-ball}) is replaced by
\begin{eqnarray}
  \label{q-ball-scaled}
 U=\mu_0^2\tilde{\psi}^2-\tilde{\lambda}\tilde{\psi}^4+\tilde{\psi}^6
 \,.
\end{eqnarray}
Following Refs.~\cite{Volkov:2002aj,Herdeiro:2021mol}, we fix
\begin{eqnarray}
  \label{scalingQcon}
    \mu_0^2=1.1
    \,, \qquad
    \tilde{\lambda}=2
    \,,
\end{eqnarray}
so that the model depends only on $\tilde{\alpha}$. For notational convenience, we henceforth drop the tildes on the dimensionless variables defined above. The coupling satisfies $0\le \alpha < \infty$. In the limit $\alpha\to 0$, gravity decouples and solitonic BSs reduce to Q-balls. In the opposite limit $\alpha\to\infty$, the quartic and sextic terms in (\ref{q-ball-scaled}) are suppressed and solitonic BSs reduce to mini-BSs. As we will show below, decreasing $\alpha$ weakens the coupling of the scalar field to gravity and significantly modifies the bifurcation structure of the solitonic BSs.

The SBSs and their zero modes are computed with the collocation solver COLSYS~\cite{Ascher:1979iha}. For a fixed coupling, the frequency $\omega$ of the background plays the role of an eigenvalue. In the following, $\omega$ at a zero mode always refers to this background frequency, since the perturbation functions in Eq.~\eqref{perturbation} do not depend on time. We solve Eqs.~\eqref{bgequ_spherical} and \eqref{p-eq-system} simultaneously as a single boundary value problem, treating $\omega$ as an additional unknown function that obeys $d\omega/dr = 0$. This raises the total differential order by one and requires one extra condition, for which we impose the normalization
\begin{equation}\label{H-normalization}
  H(r_{\rm mid}) = h_0
\end{equation}
at an intermediate radius $r_{\rm mid}$ away from the node of $H$.\footnote{The function $H$ has a single node, whose position changes along each family of solutions. If $r_{\rm mid}$ coincides with it, Eq.~\eqref{H-normalization} cannot be satisfied with $h_0 \neq 0$. If $r_{\rm mid}$ is close to it, $H(r_{\rm mid})$ is very sensitive to small changes of the solution, and the Newton iteration of the solver fails to converge.} This condition excludes the trivial solution and fixes the arbitrary amplitude of the linear perturbations; we choose $h_0$ such that $\max|H| \simeq 1$, and neither $r_{\rm mid}$ nor $h_0$ affects $\omega$. The typical relative accuracy of the solutions reported here is around $10^{-12}$.

The ABSs are obtained by solving Eqs.~\eqref{bgequ_axial} with a professional finite difference solver~\cite{Schonauer:1989zwe,SCHONAUER1990279,SCHONAUER2001473} employing a Newton-Raphson method.
To compactify the radial domain, we introduce
\begin{eqnarray}
    x=\frac{r}{c+r}\,,
\end{eqnarray}
which maps $r\in[0,\infty)$ to $x\in[0,1]$. The constant $c$ is arbitrary (we typically set $c=1$) and controls the radial grid distribution. For a fixed number of grid points, choosing $c$ appropriately improves accuracy. The equations are discretized on a grid with typically 251 points in $x$ and 50 points in $\theta$, covering $0\le x\le1$ and $0\le\theta\le\pi/2$. For all ABSs reported here, the relative difference is below $10^{-3}$.

\section{Numerical results}\label{sec:results}
\subsection{Massive BSs}
\begin{figure}[h!]
  \centering
    \includegraphics[width=0.7\textwidth]{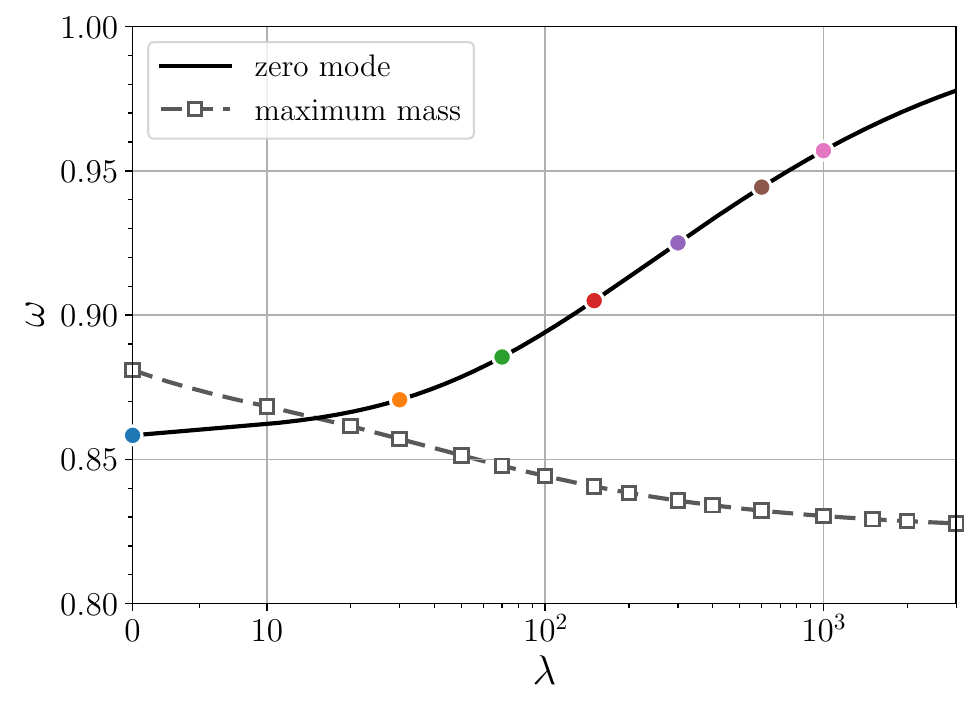}
  \caption{Frequency $\omega$ of the massive SBSs at the $\ell=2$ zero mode as a function of $\lambda$ (solid line), compared with that of the maximum-mass solution of the same family (squares, connected by a dashed fitting curve). The colored dots mark the values of $\lambda$ shown in Fig.~\ref{fig2}, with the same colors. The two curves cross at $\lambda\approx15$. For larger $\lambda$, the zero mode lies on the branch between the maximum-mass solution and the limit $\omega\to1$. The horizontal axis is linear for $\lambda<10$
and logarithmic above.
}
  \label{fig1}
\end{figure}
Sufficiently strong quartic self-interactions stabilize excited SBSs in spherically symmetric evolutions~\cite{Sanchis-Gual:2021phr,Brito:2023fwr}, whereas fully nonlinear 3+1 evolutions indicate that they do not stabilize them against generic perturbations~\cite{Brito:2025rld}. The stabilizing effect of self-interactions thus depends on the type of perturbations considered. A natural question is then how self-interactions affect the $\ell=2$ perturbations of excited SBSs, whose zero mode indicates the bifurcation into chains and rings~\cite{Liang:2025myf}. If the self-interaction in Eq.~\eqref{quartic} removed this zero mode, the bifurcation would be suppressed. To test this, we track the $\ell=2$ zero mode of the massive SBSs with one node as $\lambda$ increases.

\begin{table*}[t]
\centering
\caption{\label{tab:zeromodes}Massive SBSs at the $\ell=2$ zero mode and at the maximum mass, for the values of $\lambda$ in Fig.~\ref{fig2}. Here $\psi(0)$, $\omega$, $M$ and $Q$ denote the central value of the scalar field, the frequency, the ADM mass and the Noether charge. For $\lambda\gtrsim15$, the zero mode lies on the candidate stable branch.}
\setlength{\abovetopsep}{4pt}
\begin{tabular*}{\textwidth}{@{\extracolsep{\fill}}rcccccccc@{}}
\toprule
 & \multicolumn{4}{c}{Zero mode} & \multicolumn{4}{c}{Maximum mass} \\
\cmidrule(lr){2-5}\cmidrule(lr){6-9}
$\lambda$ & $\psi(0)$ & $\omega$ & $M$ & $Q$ & $\psi(0)$ & $\omega$ & $M$ & $Q$ \\
\midrule
0 & 0.23740 & 0.8583 & 1.328 & 1.360 & 0.17678 & 0.8810 & 1.356 & 1.392 \\
30 & 0.09929 & 0.8707 & 1.913 & 1.988 & 0.11308 & 0.8571 & 1.929 & 2.006 \\
70 & 0.05816 & 0.8855 & 2.325 & 2.421 & 0.08028 & 0.8478 & 2.471 & 2.588 \\
150 & 0.03349 & 0.9050 & 2.735 & 2.839 & 0.05630 & 0.8406 & 3.262 & 3.437 \\
300 & 0.01965 & 0.9250 & 3.052 & 3.150 & 0.04009 & 0.8357 & 4.336 & 4.588 \\
600 & 0.01131 & 0.9444 & 3.239 & 3.320 & 0.02836 & 0.8322 & 5.878 & 6.240 \\
1000 & 0.00745 & 0.9570 & 3.264 & 3.328 & 0.02193 & 0.8304 & 7.419 & 7.890 \\
\bottomrule
\end{tabular*}
\end{table*}

We find no sign that the zero mode disappears as $\lambda$ grows. The solid line in Fig.~\ref{fig1} shows the frequency of the SBS at the zero mode as a function of $\lambda$. As $\lambda$ increases, the zero mode moves to higher $\omega$, toward the Newtonian limit $\omega\to1$. We have followed it up to $\lambda\approx3000$, and at this value $\omega$ is still increasing. The squares in Fig.~\ref{fig1} mark the maximum-mass solutions, whose frequency decreases as $\lambda$ increases. Since the candidate stable branch lies between the maximum-mass solution and the Newtonian limit, it corresponds to the region between the dashed line and $\omega=1$ in this figure.

The two curves cross at $\lambda\approx15$. For $\lambda\lesssim15$, the zero mode lies beyond the maximum-mass solution, outside the candidate stable branch, whereas for $\lambda\gtrsim15$ it lies on this branch. Increasing $\lambda$ therefore does not remove the zero mode, but moves it onto the candidate stable branch. This is also seen in Table~\ref{tab:zeromodes}, where, for $\lambda\geq30$, the central value $\psi(0)$ of the zero-mode solution is smaller than that of the maximum-mass solution.
\begin{figure}[h!]
  \centering
    \includegraphics[width=1\textwidth]{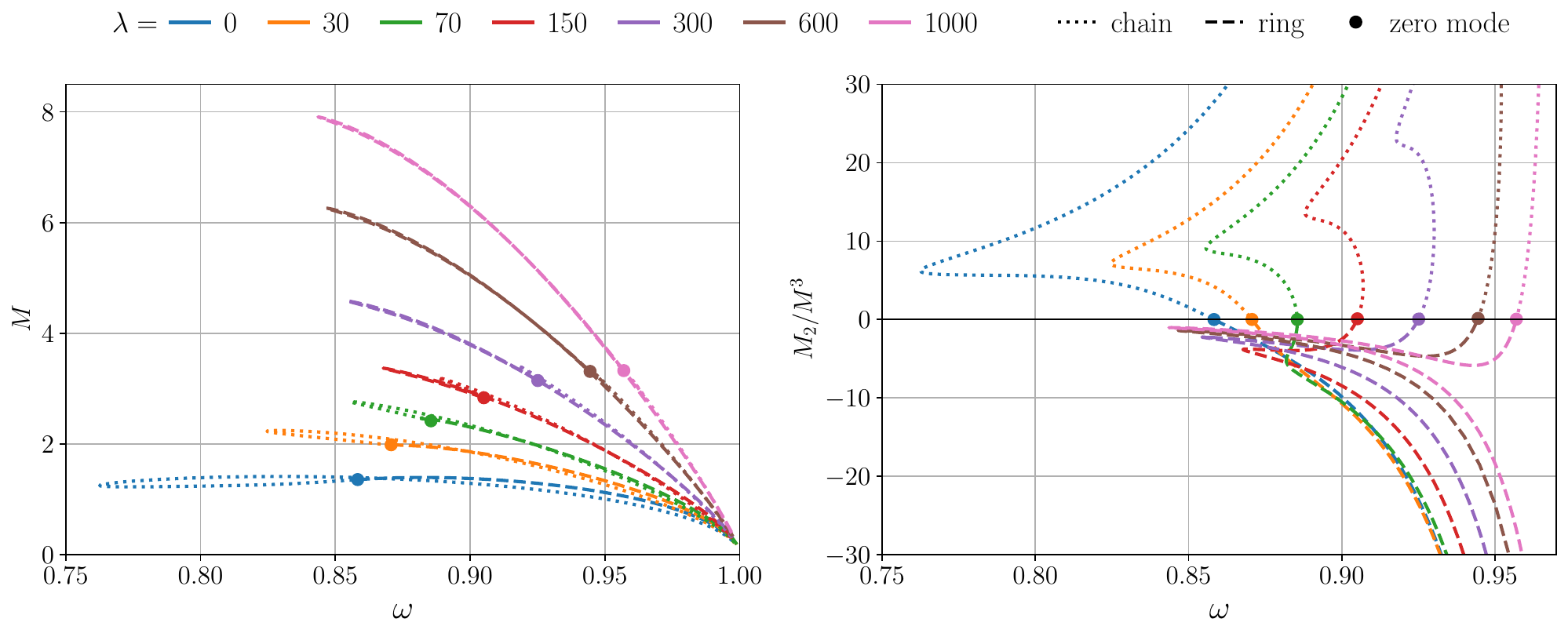}
  \caption{ADM mass $M$ (left) and normalized quadrupole moment $M_2/M^3$ (right) as functions of $\omega$ for the chains (dotted lines) and rings (dashed lines) of massive BSs, for several values of $\lambda$. The dots mark the $\ell=2$ zero modes of the SBSs listed in Table~\ref{tab:zeromodes}, where the chains and rings bifurcate from the SBSs.}
  \label{fig2}
\end{figure}
Since, generically, the SBSs on the two sides of a zero mode differ by one unstable $\ell=2$ mode, for $\lambda\gtrsim15$, the candidate stable branch cannot be entirely stable against $\ell=2$ perturbations. This linear result is consistent with the non-spherical instability of excited SBSs found in Ref.~\cite{Brito:2025rld}, although the unstable mode seen in those evolutions need not be the $\ell=2$ mode considered here.

\newlength{\figpw}\setlength{\figpw}{0.28\textwidth}   
\newlength{\figgut}\setlength{\figgut}{0.1\figpw}      
\newlength{\figsep}\setlength{\figsep}{0.012\textwidth} 
\begin{figure}[p]
  \centering
  \hspace*{\figgut}\includegraphics[width=1.55\figpw]{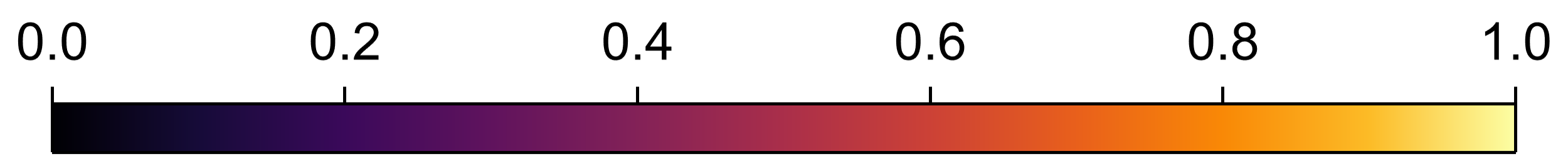}\\[2pt]
  \hspace*{\figgut}%
  \includegraphics[width=\figpw]{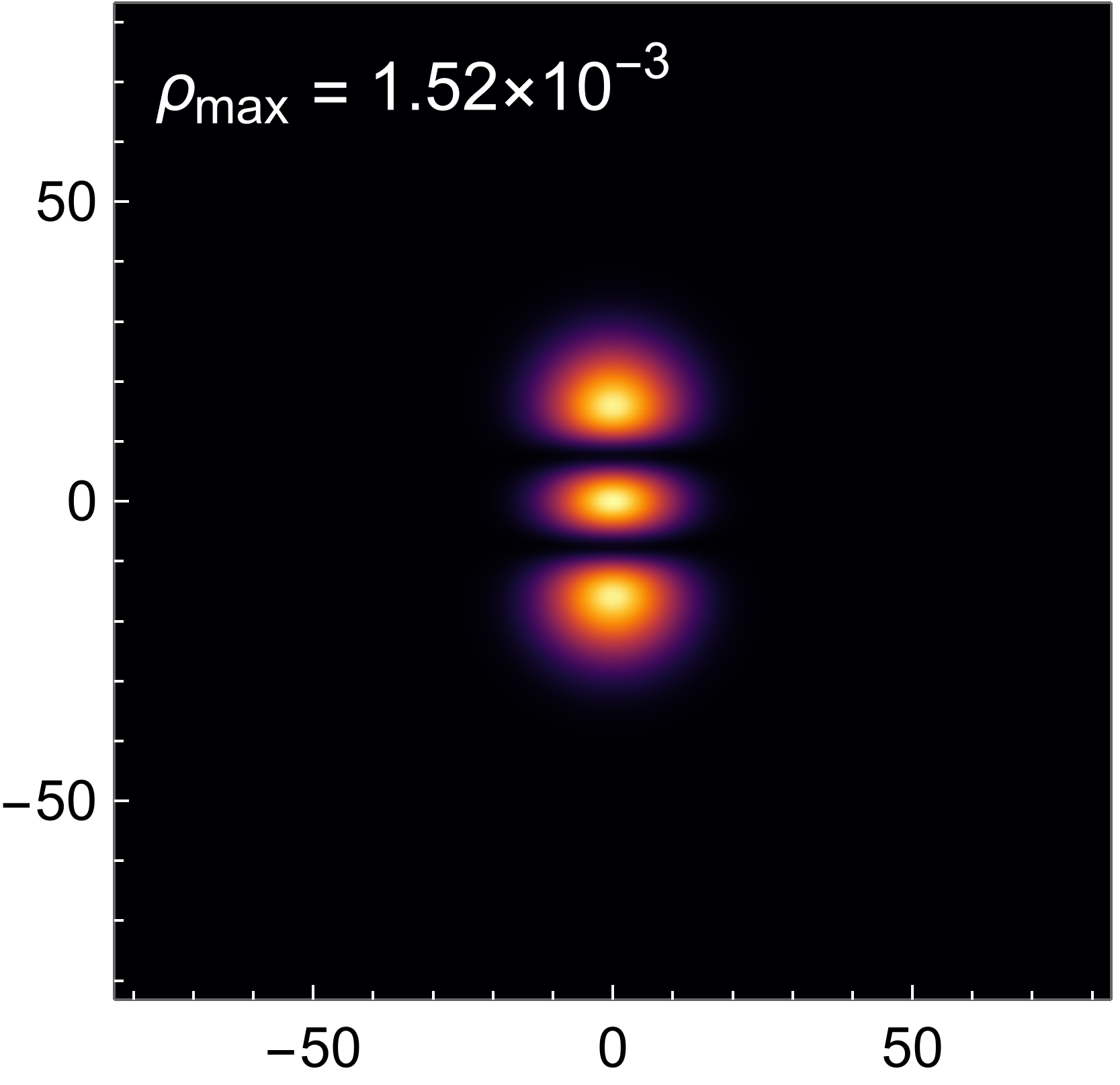}\hspace{\figsep}%
  \includegraphics[width=\figpw]{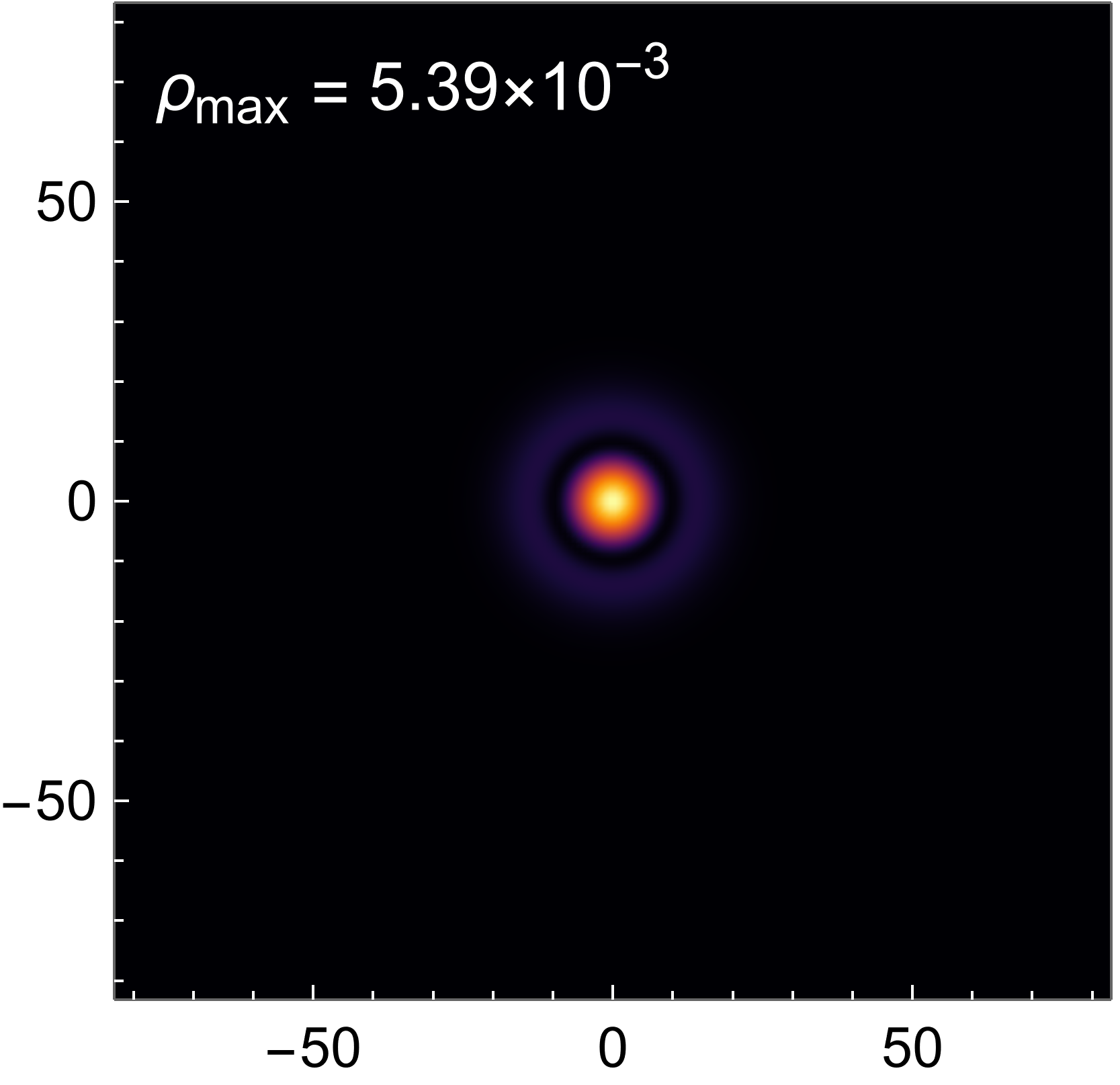}\hspace{\figsep}%
  \includegraphics[width=\figpw]{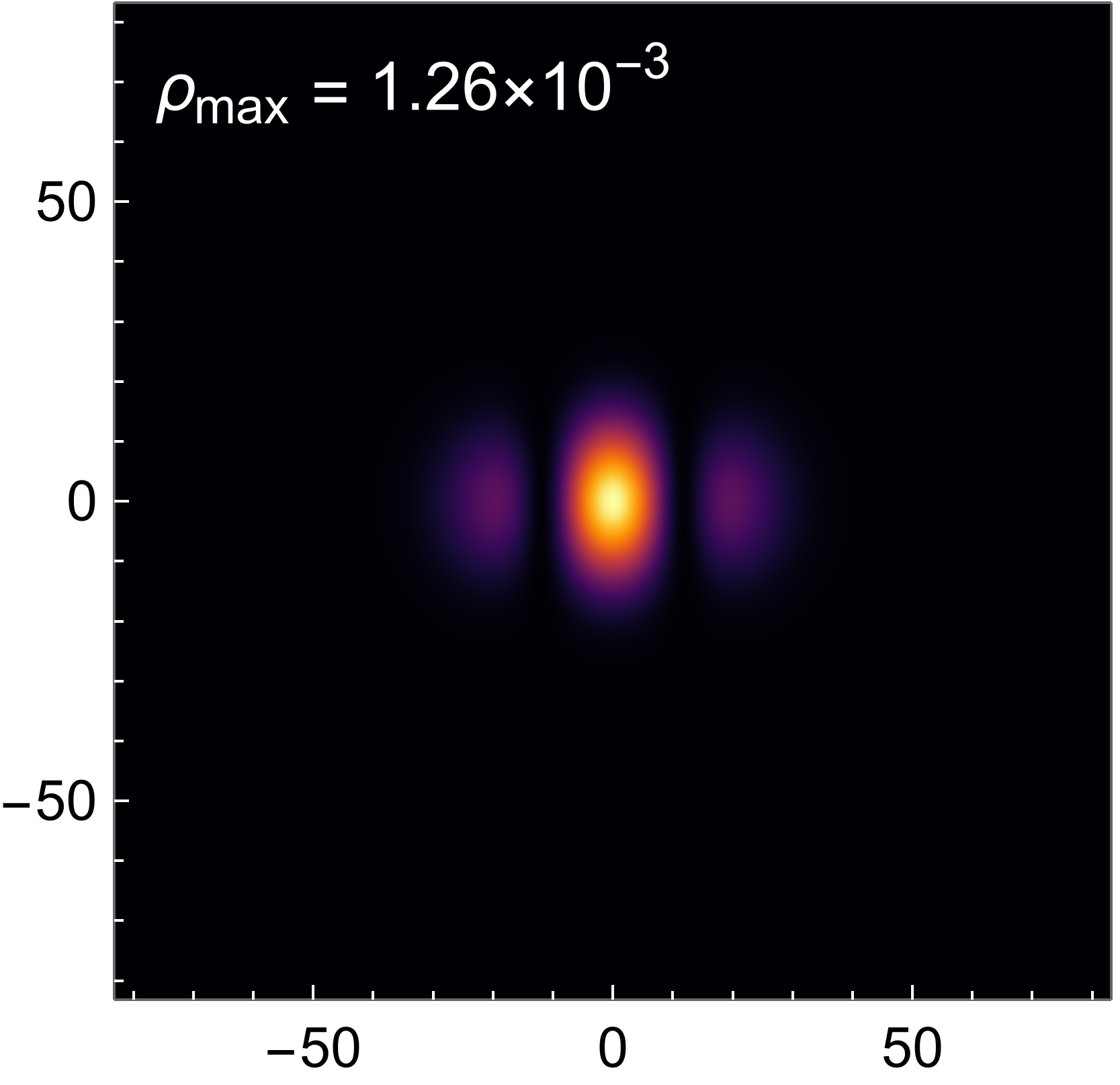}\\[1pt]
  \hspace*{\figgut}%
  \includegraphics[width=\figpw]{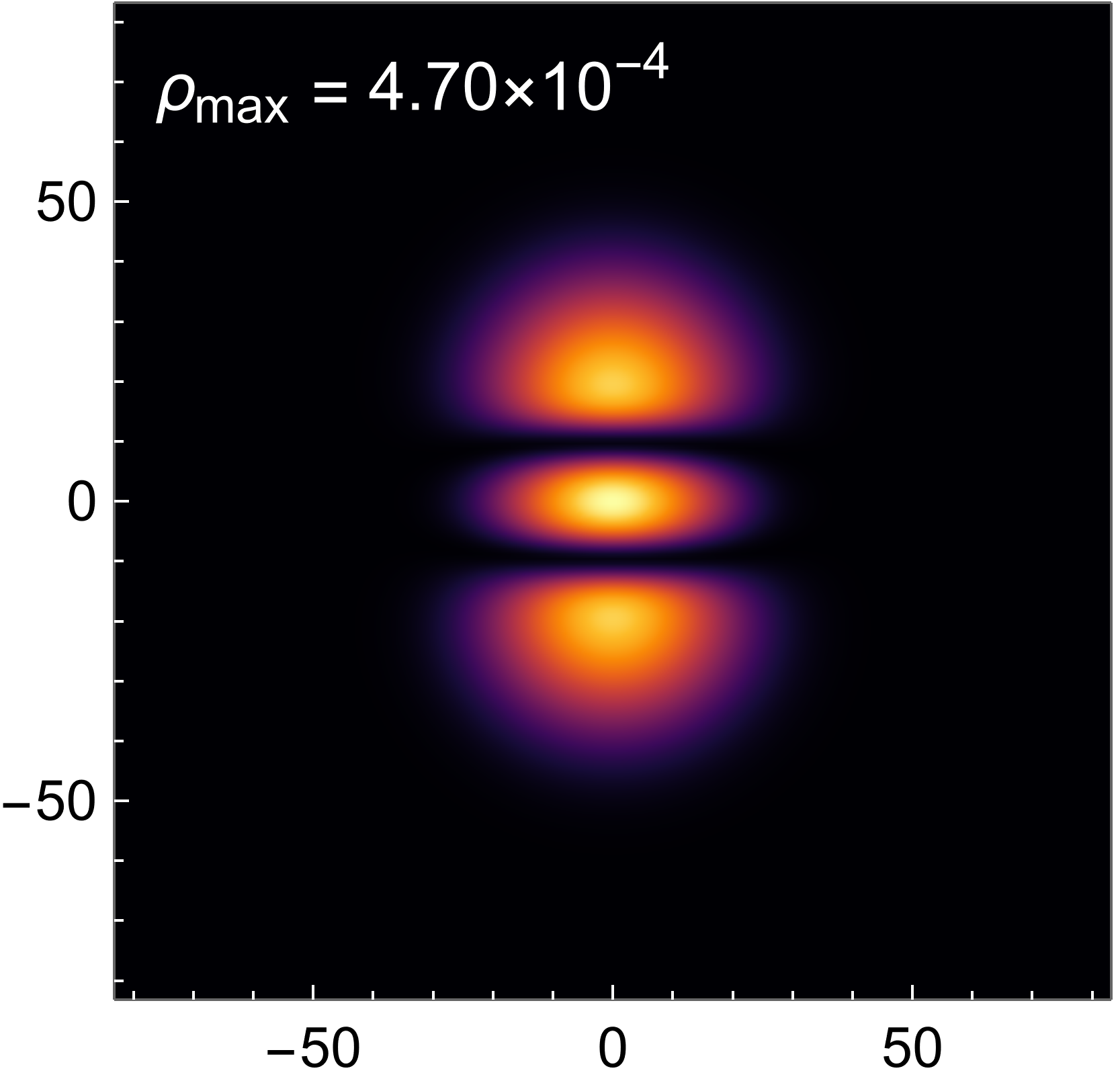}\hspace{\figsep}%
  \includegraphics[width=\figpw]{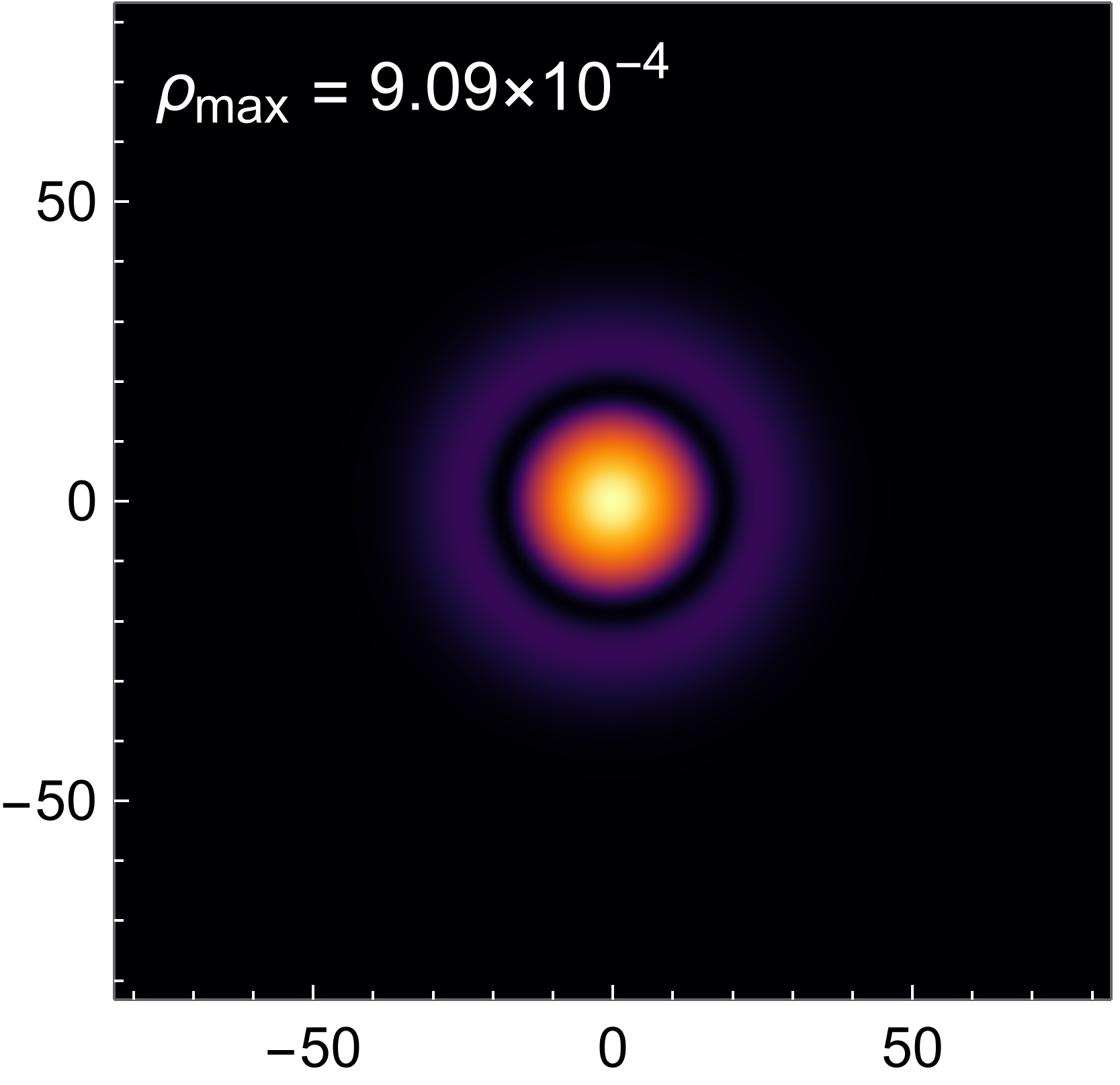}\hspace{\figsep}%
  \includegraphics[width=\figpw]{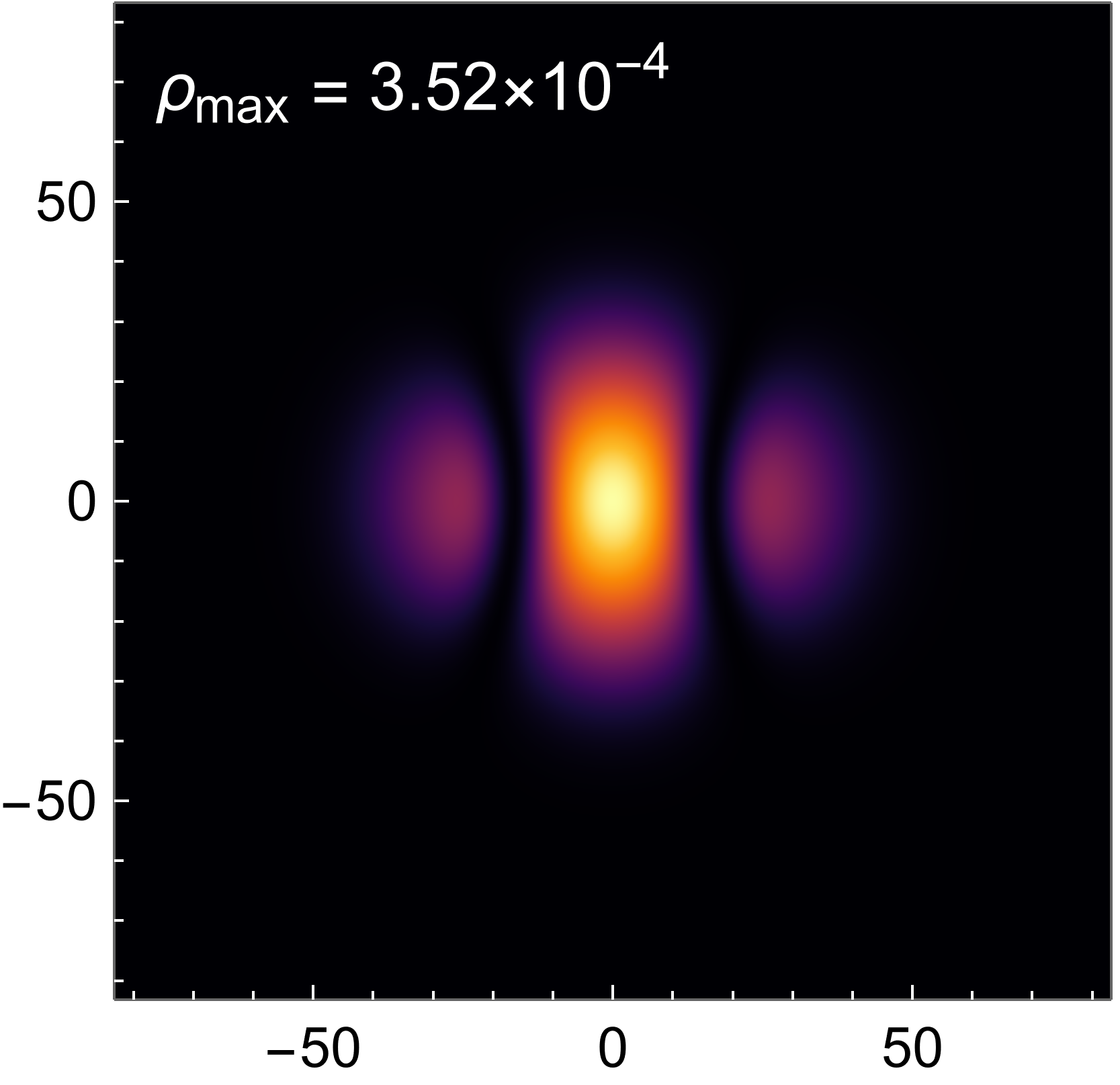}\\[1pt]
  \hspace*{\figgut}%
  \includegraphics[width=\figpw]{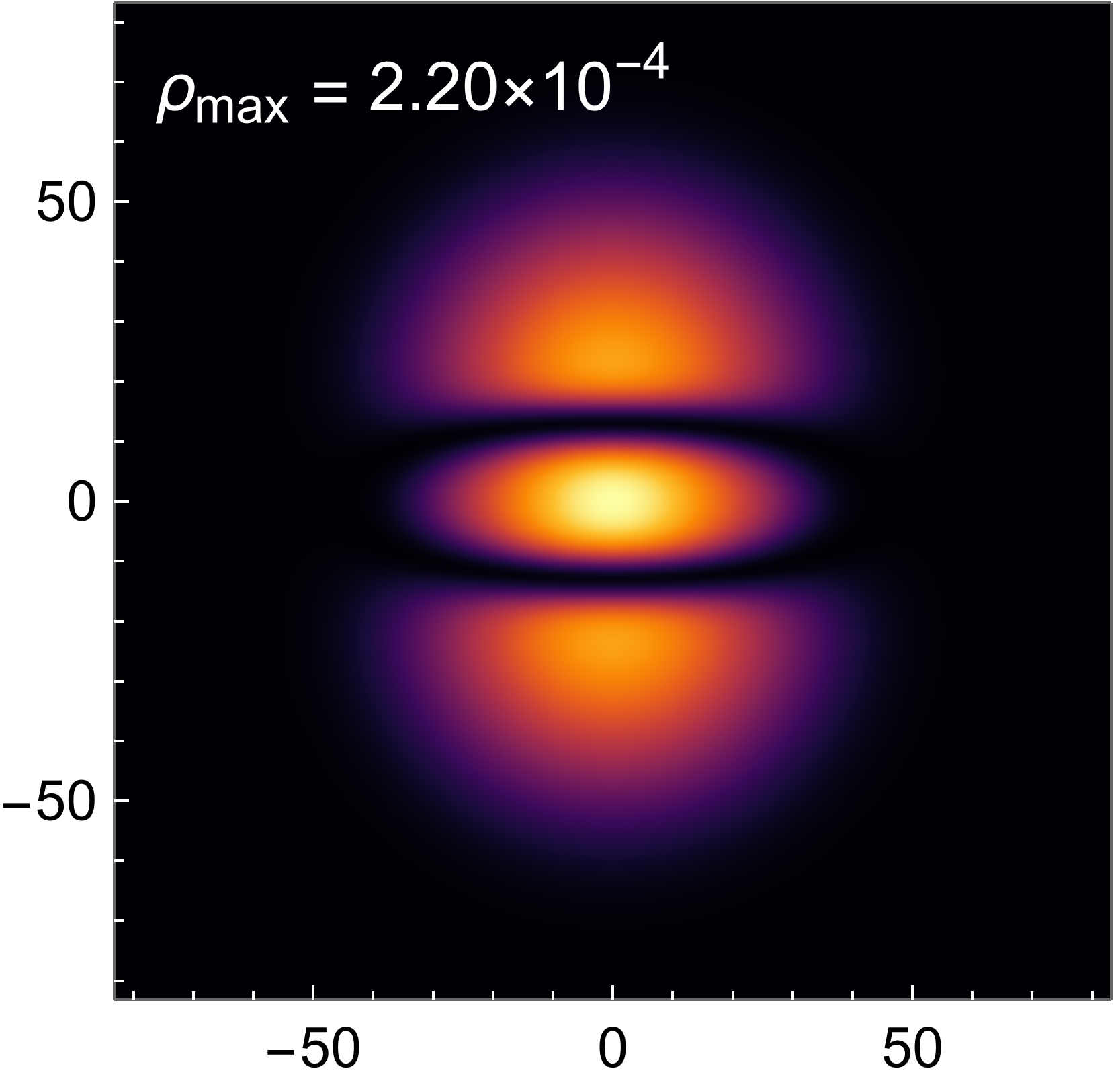}\hspace{\figsep}%
  \includegraphics[width=\figpw]{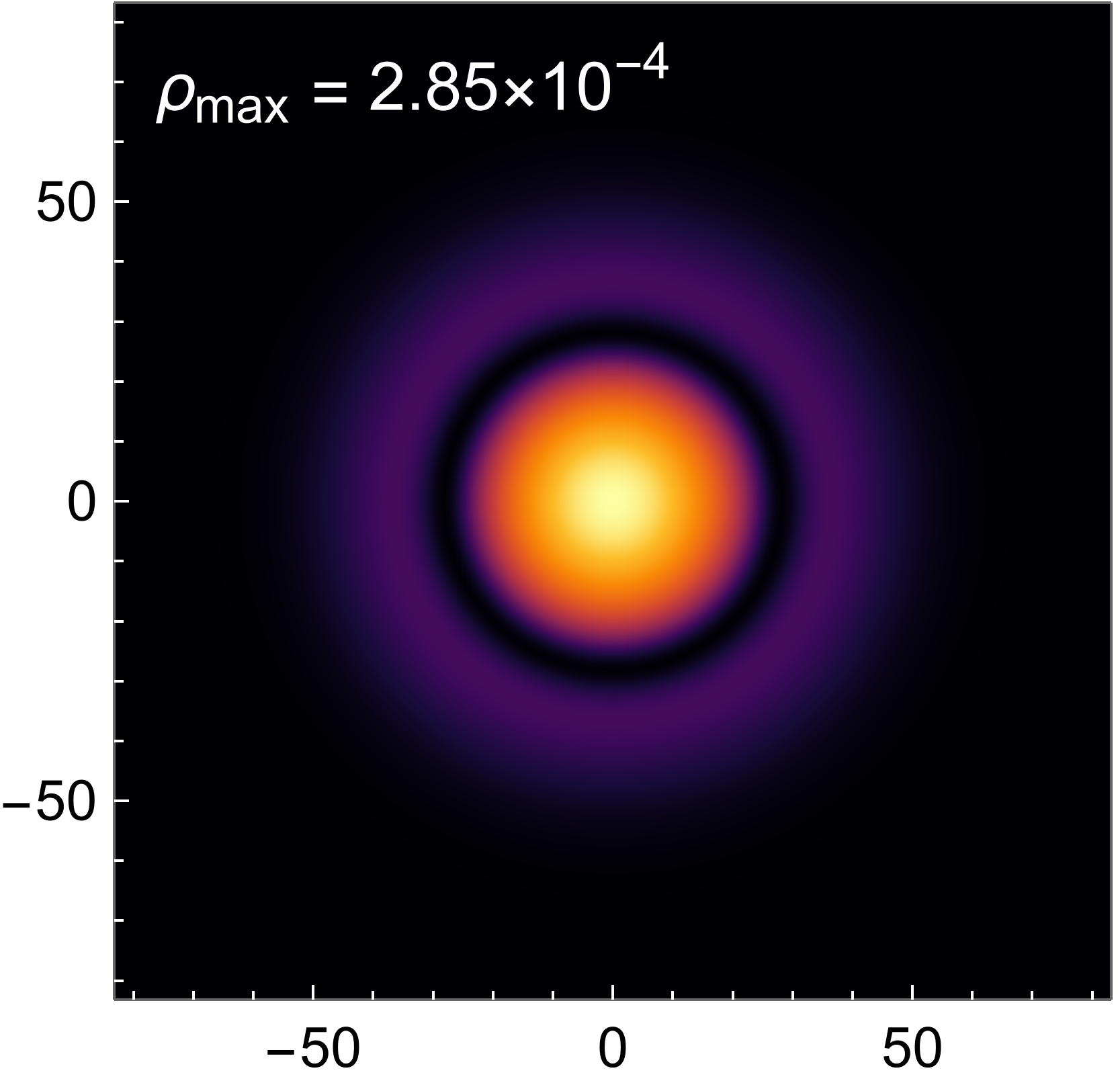}\hspace{\figsep}%
  \includegraphics[width=\figpw]{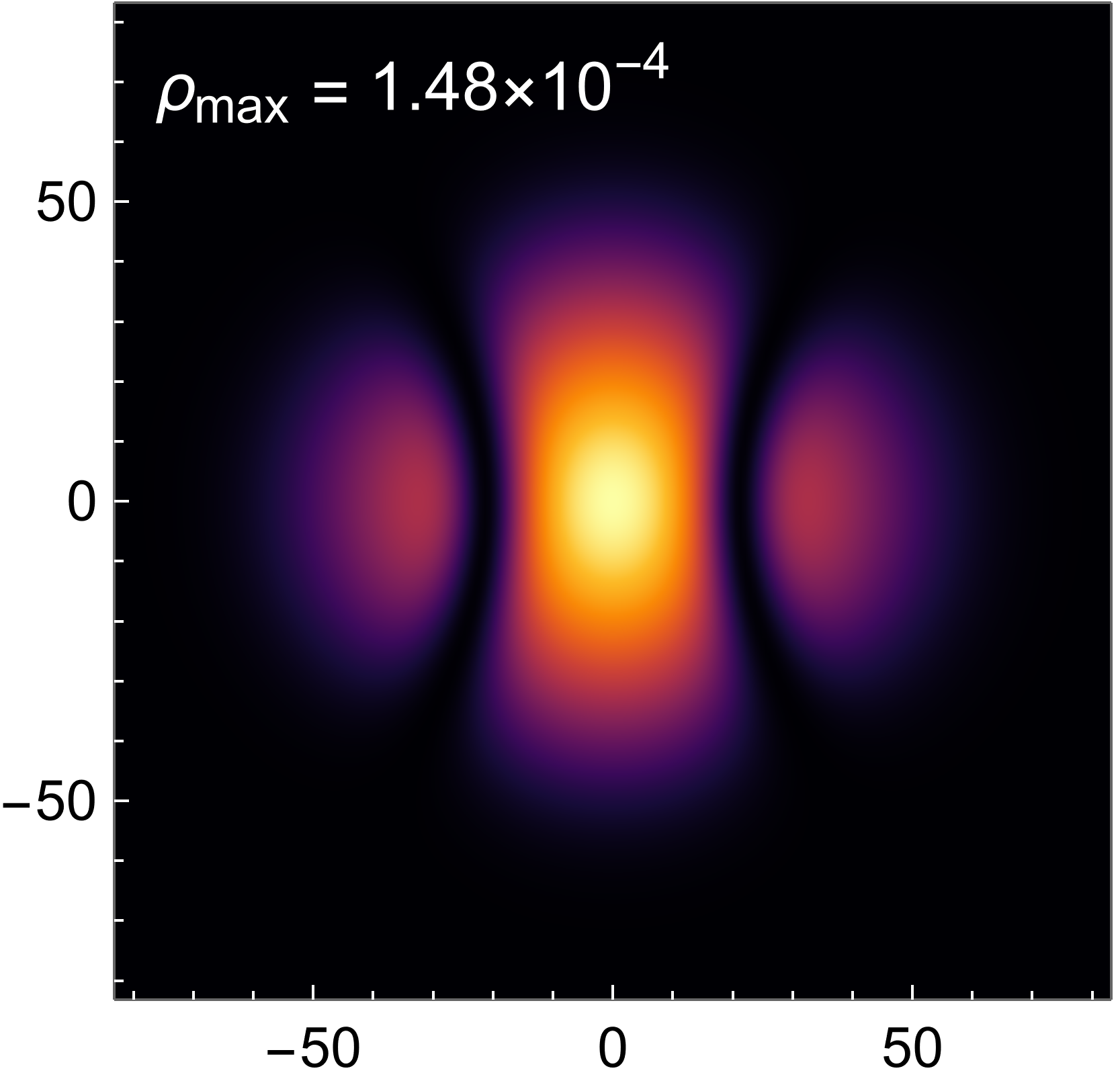}\\[4pt]
  \includegraphics[width=\dimexpr\figpw+\figgut\relax]{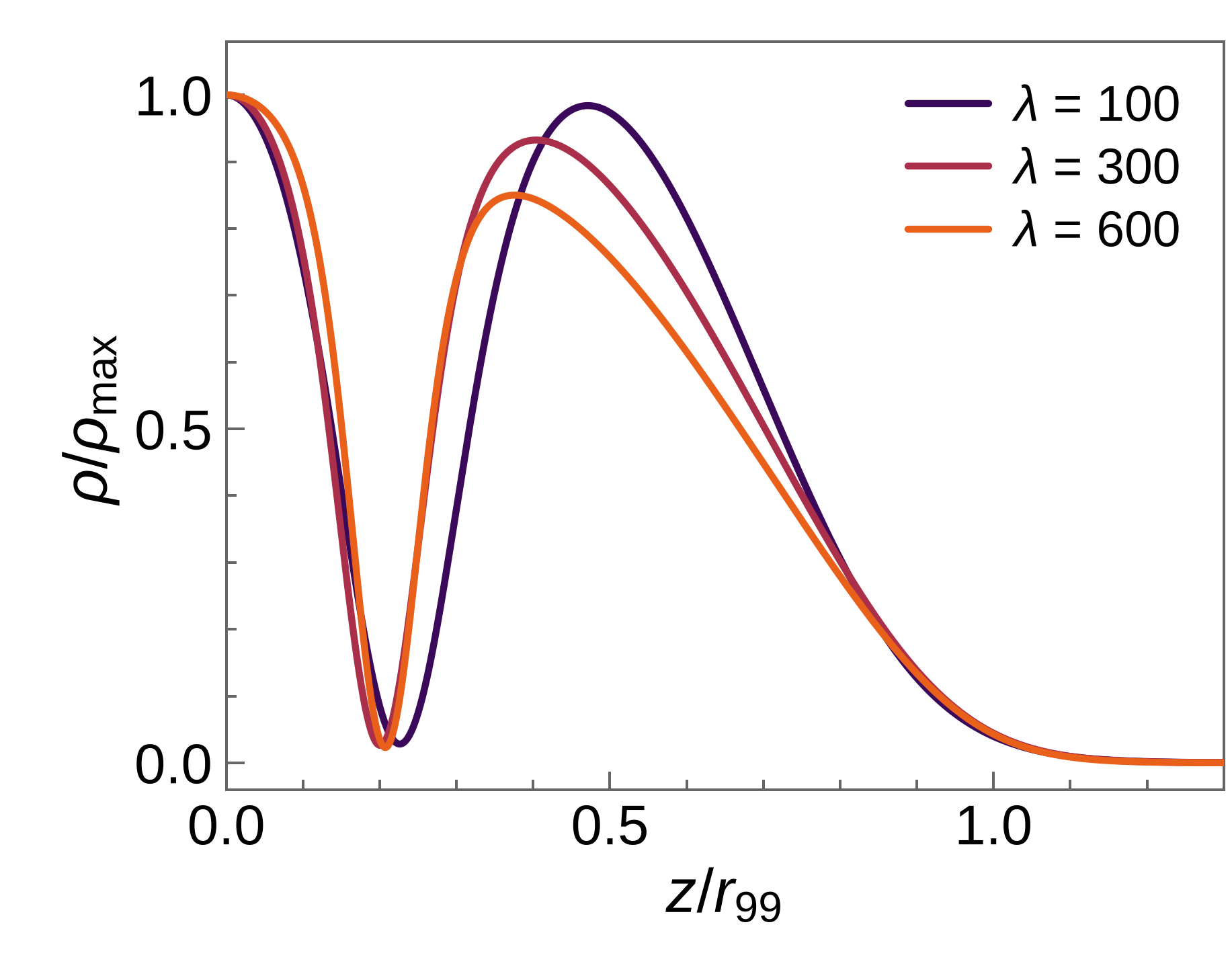}\hspace{\figsep}%
  \includegraphics[width=\figpw]{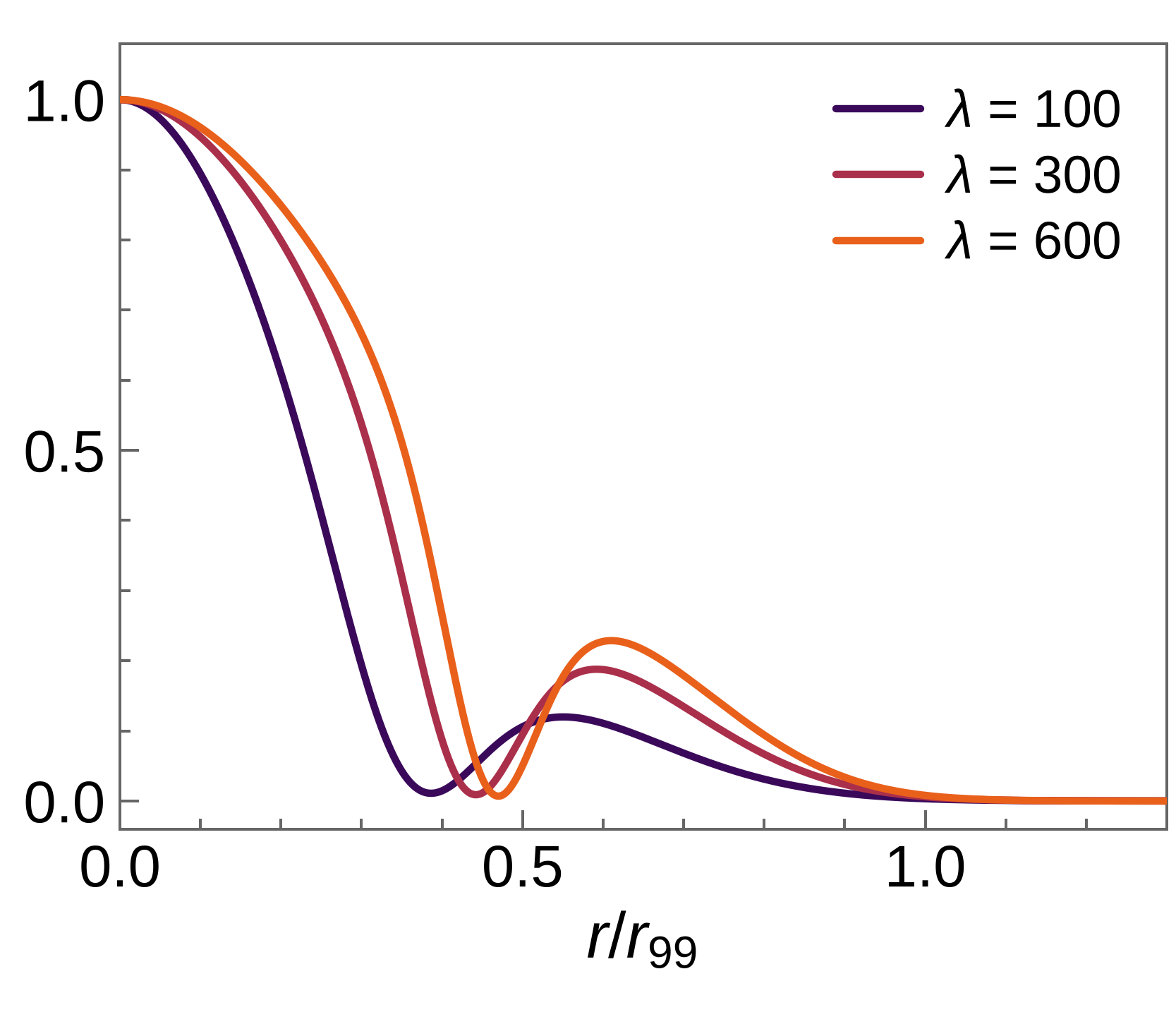}\hspace{\figsep}%
  \includegraphics[width=\figpw]{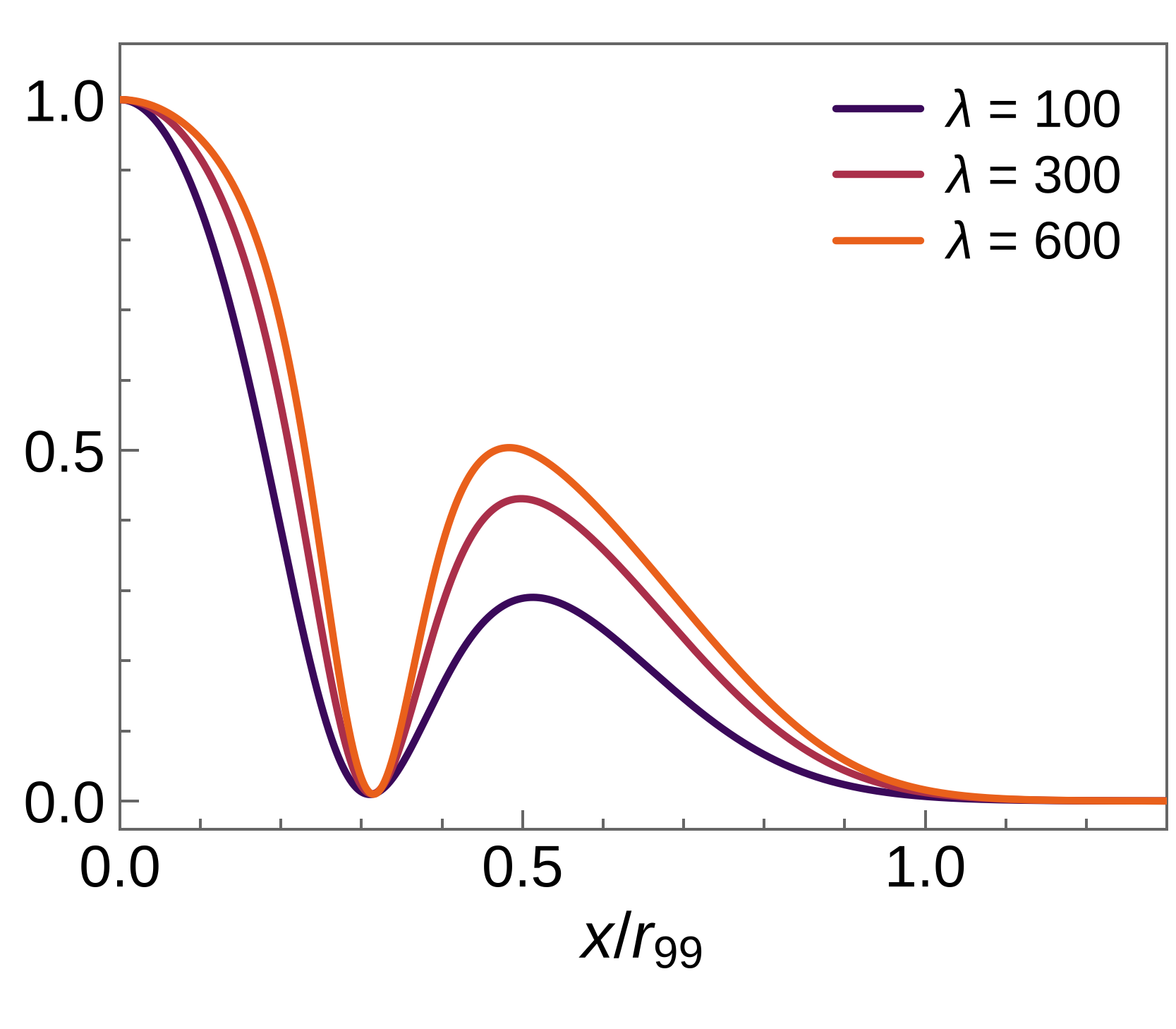}
  \caption{Normalized energy density $\rho/\rho_{\max}$ in the $(x,z)$ plane, with $(x,z)=(r\sin\theta,\,r\cos\theta)$ and $r,\theta$ as in Eq.~\eqref{eq1.6}, for chains with $M_2/M^3=50$ (left), SBSs at the zero mode (middle) and rings with $M_2/M^3=-50$ (right). The first to third rows correspond to $\lambda=100$, 300 and 600, and the symmetry axis $z$ is vertical. The bottom row shows $\rho/\rho_{\max}$ of the same chains, SBSs and rings along the $z$ axis, the radial direction and the $x$ axis, respectively, with lengths in units of the corresponding effective coordinate radius $r_{99}$.}
  \label{fig3}
\end{figure}
To confirm that the zero modes obtained from the linearized equations mark bifurcations, and to follow the bifurcating solutions into the nonlinear regime, we have constructed the corresponding chains and rings for the values of $\lambda$ marked by the colored dots in Fig.~\ref{fig1}. The results are shown in Fig.~\ref{fig2}. The left panel shows the $(\omega,M)$ diagram of these ABSs. For each $\lambda$, the chain and ring branches both extend to the Newtonian limit $\omega\to1$ and meet at the zero mode, so that they form a loop. This can be clearly seen for small $\lambda$. As $\lambda$ increases, the frequency range of the ABSs first shrinks and then grows, while their maximum mass increases monotonically, like that of the SBSs. At the same time, the loop becomes narrower, and for large values of $\lambda$, its two sides almost coincide.

The right panel shows more clearly how the two branches are connected to the zero mode. Since $M_2$ vanishes for the SBSs, the two branches meet at $M_2=0$, at the zero modes marked by the dots, with $M_2>0$ for the chains and $M_2<0$ for the rings. This panel also shows that the minimum frequency of the ABSs is reached on the chain branch for small $\lambda$, but on the ring branch for large $\lambda$.

The quadrupole moment measures the deformation of the ABSs by a single number. To see how their energy is distributed, Fig.~\ref{fig3} shows the energy density $\rho$, normalized to its maximum $\rho_{\max}$, of chains and rings with $M_2/M^3=\pm50$ and of the SBSs at the zero mode, for $\lambda=100$, 300 and 600. The SBSs have the structure of a central core surrounded by a spherical shell, separated by a nodal surface of the scalar field on which $\rho$ nearly vanishes. In the chains, the outer part is concentrated near the symmetry axis and forms two components above and below the core. In the rings, it is concentrated near the equatorial plane and forms a torus around the core.

In both cases, the core is deformed in the opposite direction to the outer part, being flattened along the axis in the chains and elongated along it in the rings. This is because the core and the outer part share the nodal surface of $\psi$ as their boundary, so that where this surface moves inward, the core shrinks and the outer part extends inward. In the chains, the nodal surface moves inward near the symmetry axis, and in the rings near the equatorial plane. The deformation of the core is not a tidal effect of the outer part, which would act the other way. As a result of the opposite deformations, the contribution of the core to $M_2$ partly cancels that of the outer part.

With increasing $\lambda$, all configurations become larger while $\rho_{\max}$ decreases, since the repulsive self-interaction makes the stars more dilute. To compare their shapes, the bottom row of Fig.~\ref{fig3} shows the profiles in units of $r_{99}$, the coordinate radius within which 99\% of the Komar mass in Eq.~\eqref{eq:M} is contained. In these units, the node of the SBSs moves outward as $\lambda$ increases. Moreover, the outer maximum of $\rho/\rho_{\max}$, which belongs to the outer part, increases with $\lambda$ for the SBSs and the rings, whereas it decreases for the chains.

\subsection{Solitonic BSs}
\begin{figure}[h!]
  \centering
    \includegraphics[width=0.49\textwidth]{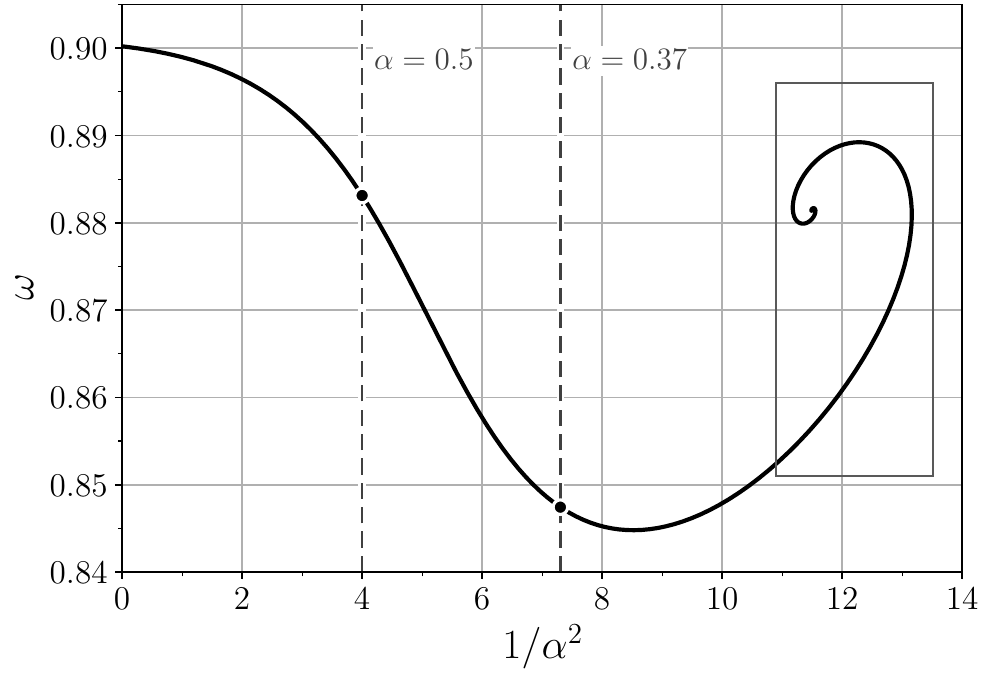}
    \hfill
    \includegraphics[width=0.49\textwidth]{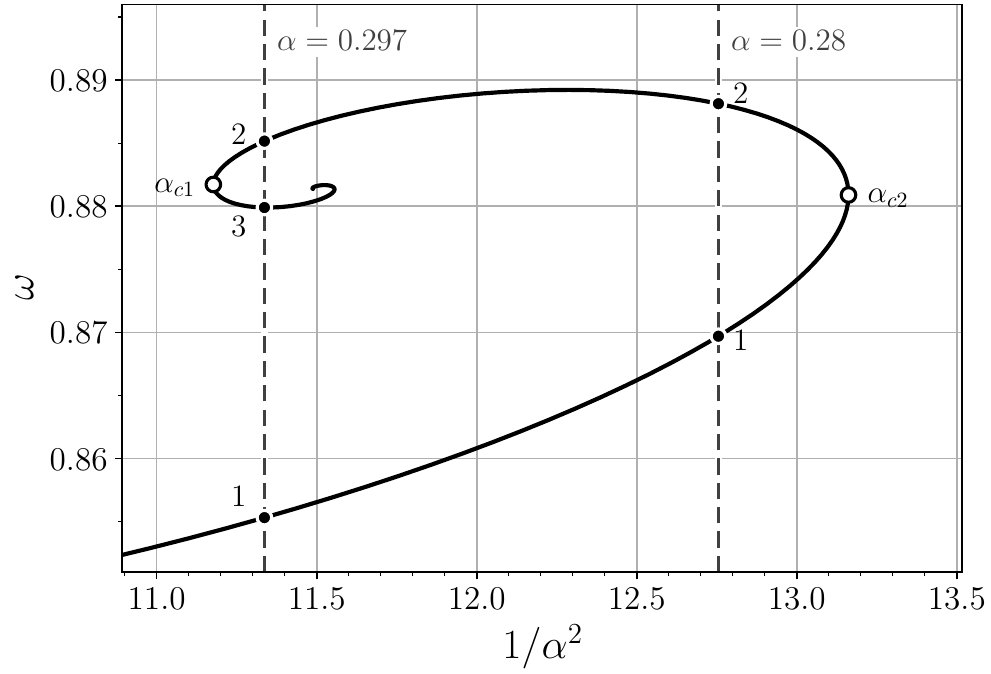}
  \caption{Frequency $\omega$ of the SBSs at the $\ell=2$ zero modes as a function of $1/\alpha^2$ (left). The right panel zooms into the boxed region. As $\alpha\to\infty$ ($1/\alpha^2=0$), the curve ends at the zero mode of the first excited mini-BS, $\omega\simeq0.9002$. The dashed lines mark $\alpha=0.5$ and $0.37$, where the SBSs have a single zero mode, and $\alpha=0.28$ and $0.297$, where they have two and three zero modes, respectively. The black dots represent these zero modes. The open circles mark the critical values $\alpha_{c1}\simeq0.2991$ and $\alpha_{c2}\simeq0.2756$. Zero modes exist only for $\alpha\geq\alpha_{c2}$. For $\alpha_{c2}<\alpha<\alpha_{c1}$, a single SBS family has multiple zero modes, and hence multiple bifurcation points.}
  \label{fig:zm_alpha}
\end{figure}

As for the massive BSs, we start by tracking the zero modes along the SBS family. With the coupling constants fixed by Eq.~\eqref{scalingQcon}, we scan the gravitational coupling $\alpha$, and find a behavior quite different from that of the massive BSs. As shown in Fig.~\ref{fig:zm_alpha}, the frequency of the SBS at the zero mode does not change monotonically as $\alpha$ decreases. Instead, the curve forms a spiral at small $\alpha$. In the limit $\alpha\to\infty$, where the solitonic BSs reduce to mini-BSs, it approaches $\omega\simeq0.9002$, which is the value $\omega=0.8583$ in Table~\ref{tab:zeromodes} multiplied by $\mu_0=\sqrt{1.1}$ because of the scaling~(\ref{scalingQ}). For $\alpha>\alpha_{c1}$, there is a single zero mode at each $\alpha$. For $\alpha_{c2}<\alpha<\alpha_{c1}$, which is the range covered by the spiral, the frequency takes several values at fixed $\alpha$. In other words, a single SBS family has multiple zero modes, and hence multiple bifurcations into chains and rings. Close to the center of the spiral, there are even more zero modes, but in this work we consider at most three. For $\alpha<\alpha_{c2}$, the zero modes disappear, which shows that the $\ell=2$ zero modes of solitonic BSs exist only if gravity is strong enough.

In the following, we analyze in more detail the bifurcations of solitonic BSs at the values of $\alpha$ marked by the dashed lines in Fig.~\ref{fig:zm_alpha}. The frequency, mass, Noether charge, binding energy and $\psi(0)$ of the SBSs at these zero modes are listed in Table~\ref{tab:zm_solitonic}. The table shows that the binding energy $E_{\rm b}=M-\mu_0Q$ increases as $\alpha$ decreases, and that among these SBSs only the one at $\alpha=0.5$ has $E_{\rm b}<0$ and is therefore bound.

\begin{table}[h!]
\centering
\caption{Frequency $\omega$, mass $M$, Noether charge $Q$, binding energy $E_{\rm b}=M-\mu_0Q$, and $\psi(0)$ of the SBSs at the zero modes marked in Fig.~\ref{fig:zm_alpha}. The mass, Noether charge and binding energy are multiplied by $\alpha^2$.}\vspace{2mm}
\label{tab:zm_solitonic}
\begin{tabular}{ccccccc}
\toprule
$\alpha$ & Zero mode & $\omega$ & $M\alpha^2$ & $Q\alpha^2$ & $E_{\rm b}\alpha^2$ & $\psi(0)$ \\
\midrule
0.5   & 1 & 0.8831 & 13.352 & 12.865 & $-0.140$ & 0.7214 \\
\midrule
0.37  & 1 & 0.8474 & 10.903 & 10.178 & 0.229 & 1.2987 \\
\midrule
\multirow{3}{*}{0.297} & 1 & 0.8553 & 9.147 & 8.182 & 0.566 & 1.7226 \\
      & 2 & 0.8852 & 9.335 & 8.379 & 0.547 & 2.8172 \\
      & 3 & 0.8799 & 9.300 & 8.339 & 0.553 & 3.5192 \\
\midrule
\multirow{2}{*}{0.28} & 1 & 0.8697 & 8.765 & 7.746 & 0.641 & 1.9187 \\
      & 2 & 0.8881 & 8.870 & 7.863 & 0.623 & 2.3068 \\
\bottomrule
\end{tabular}
\end{table}

\begin{figure}[t]
  \centering
  \includegraphics[width=0.49\textwidth]{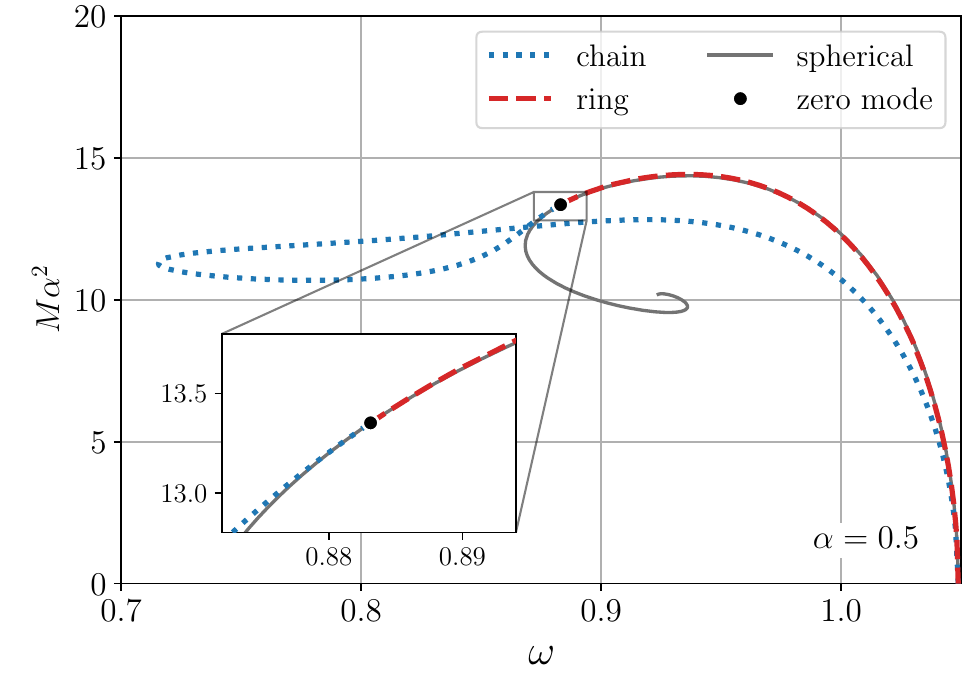}\hfill
  \includegraphics[width=0.49\textwidth]{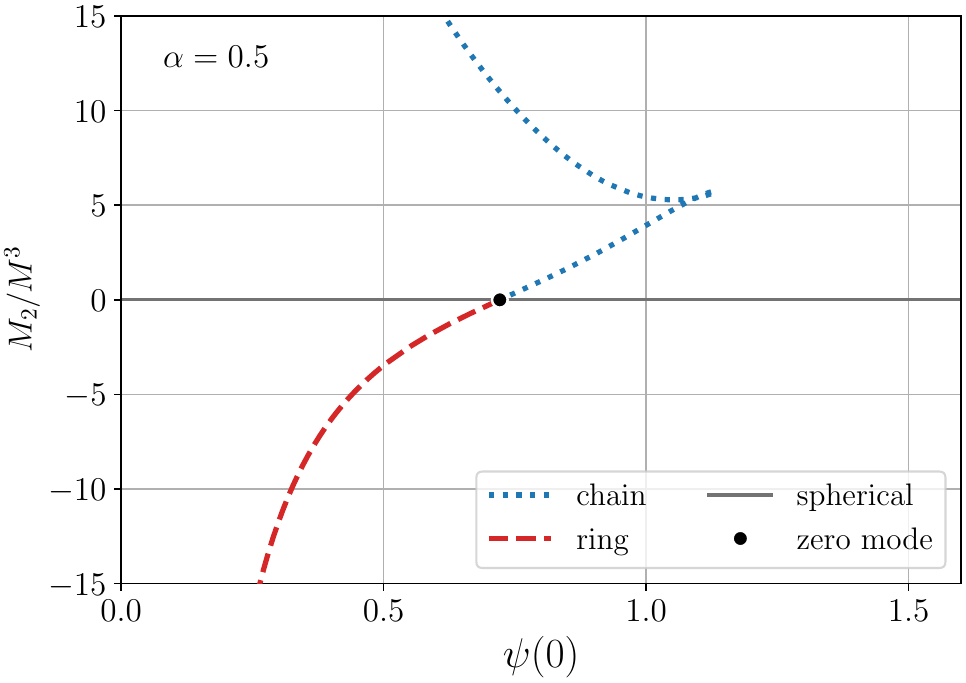}\\[4pt]
  \includegraphics[width=0.49\textwidth]{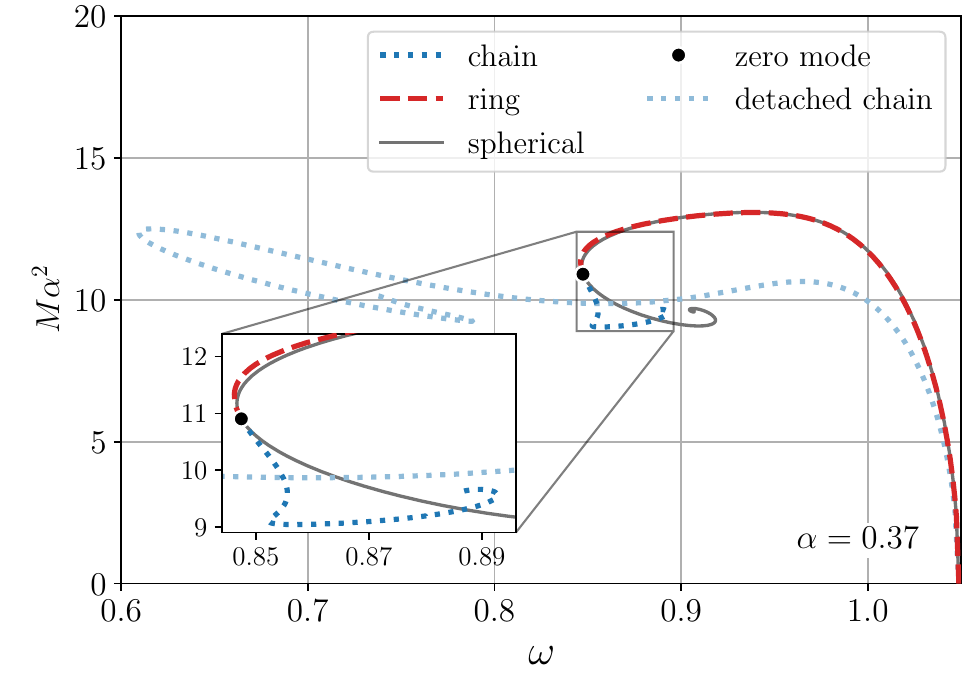}\hfill
  \includegraphics[width=0.49\textwidth]{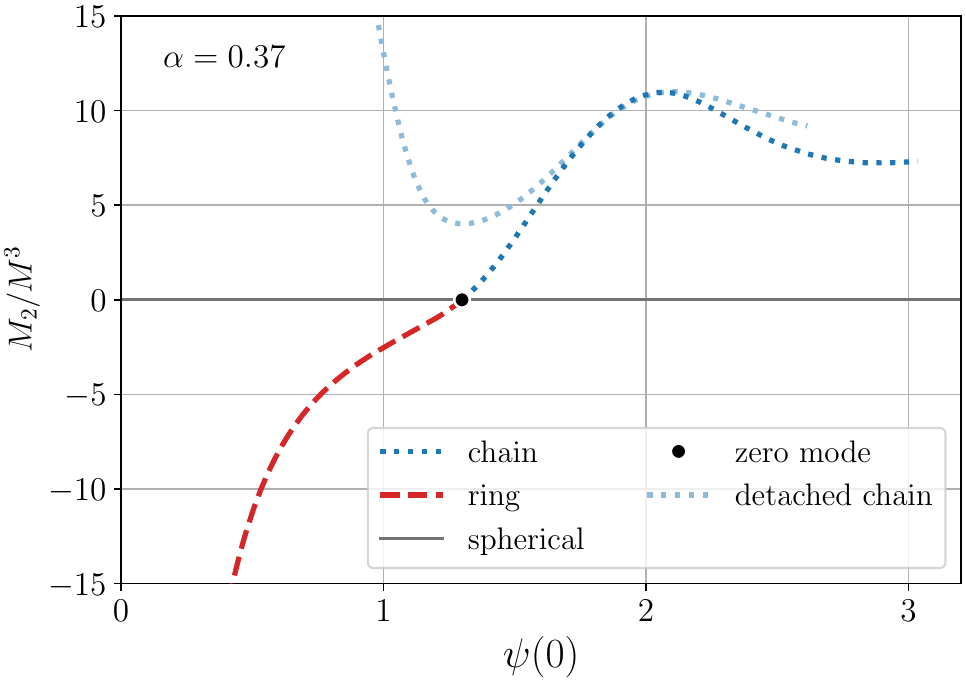}
  \caption{Scaled mass $M\alpha^2$ of the chains, rings and SBSs as a function of $\omega$ (left), and normalized quadrupole moment $M_2/M^3$ near the zero mode as a function of $\psi(0)$ (right), for $\alpha=0.5$ (top) and $\alpha=0.37$ (bottom). The insets zoom into the regions around the zero modes, where the chains and rings bifurcate from the SBSs. In both cases, the SBS family has only one zero mode. The detached chain denotes the chain branch from the Newtonian limit that is not connected to the zero mode.}
  \label{fig:single_zm}
\end{figure}

We first consider the case of a single zero mode. Figure~\ref{fig:single_zm} shows the mass and the quadrupole moment of the solitonic BSs for $\alpha=0.5$ and $0.37$. For $\alpha=0.5$, the situation is similar to that of the massive BSs. The chain and ring branches both approach the Newtonian limit $M\to0$ as $\omega$ approaches its maximum value $\mu_0$, and they meet at the zero mode, so that together they form a loop.

As $\alpha$ decreases, we find that the chain branch extending to the Newtonian limit detaches from the zero mode at $\alpha\approx 0.38$. Below this value, it starts to form a spiral instead, similar to that of the SBSs, as shown by the \emph{detached chain} in the bottom panels. However, a chain branch and a ring branch still bifurcate from the zero mode. The ring branch still extends to the Newtonian limit, but the chain branch now moves away from it. Starting from the zero mode, $\psi(0)$ increases monotonically along this chain branch, as shown in the bottom right panel. Comparing the right panels for $\alpha=0.5$ and $0.37$, we see that the single chain branch at $\alpha=0.5$ has split into two, the detached chain and the chain branch that meets the ring branch at the zero mode. We followed the latter up to $\psi(0)\approx3$. As we will see below, once the SBS family has more than one zero mode, the chain branch from the first zero mode ends at the second one.

\begin{figure}[t]
  \centering
  \includegraphics[width=0.49\textwidth]{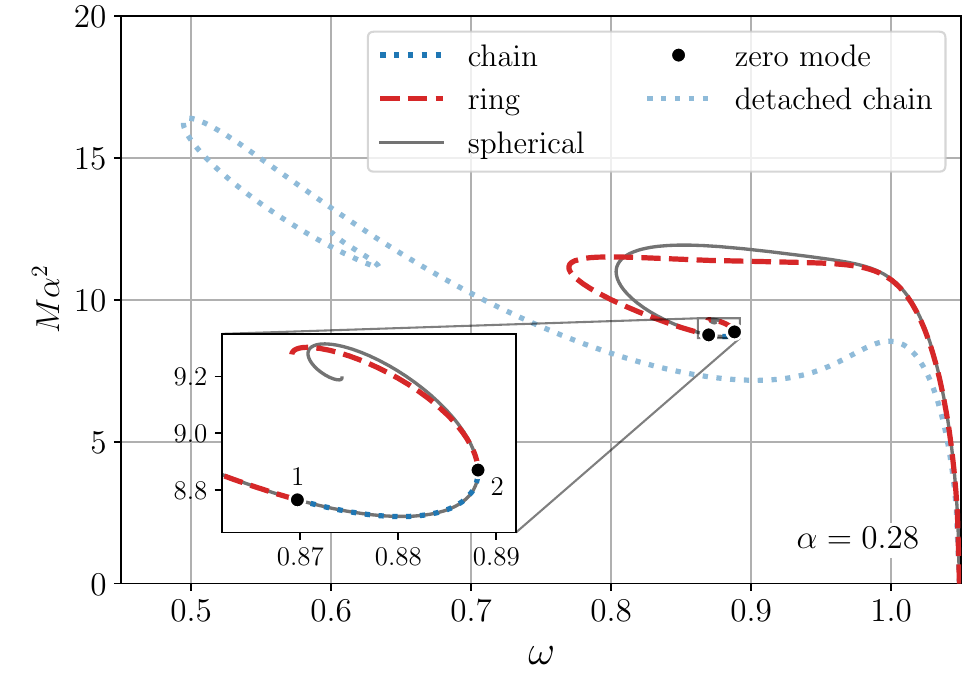}\hfill
  \includegraphics[width=0.49\textwidth]{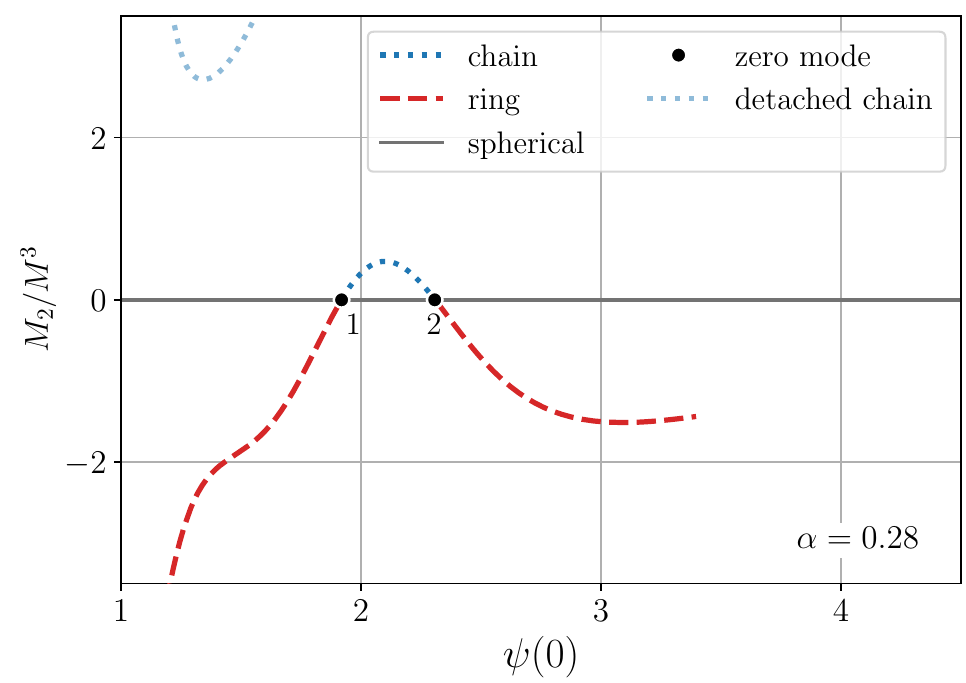}\\[4pt]
  \includegraphics[width=0.49\textwidth]{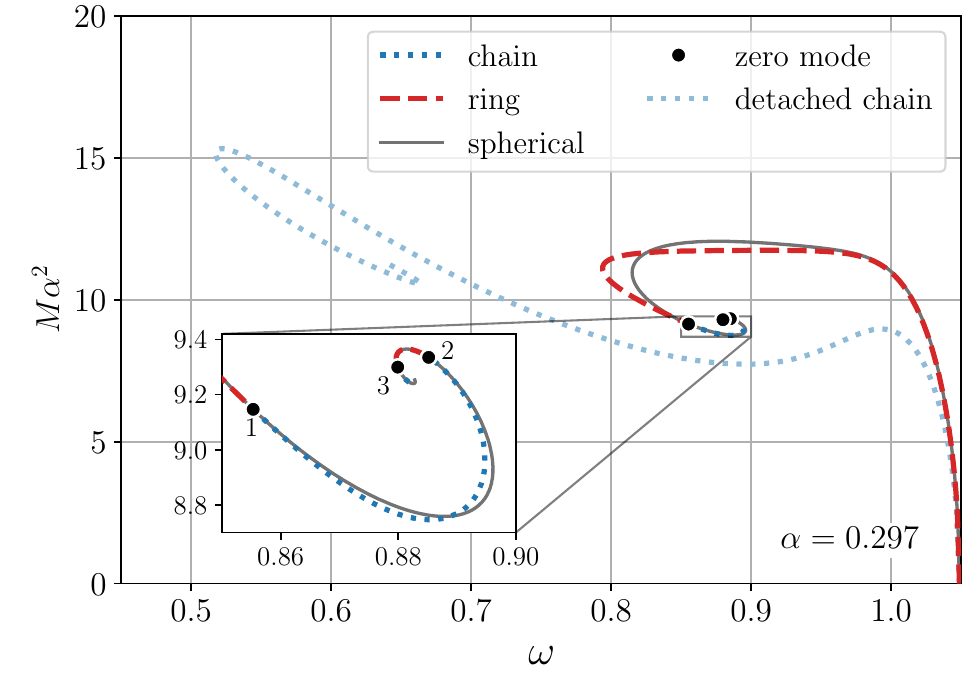}\hfill
  \includegraphics[width=0.49\textwidth]{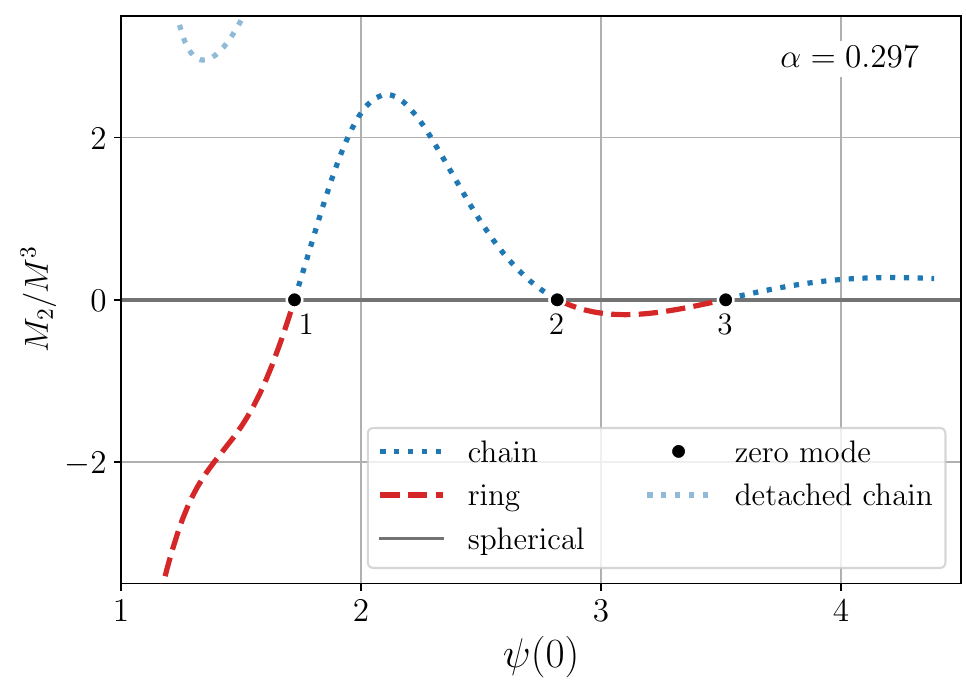}
  \caption{Same as Fig.~\ref{fig:single_zm}, but for $\alpha=0.28$ (top) and $\alpha=0.297$ (bottom), where the SBS family has two and three zero modes, respectively. The zero modes are numbered as in Table~\ref{tab:zm_solitonic}.}
  \label{fig:multi_zm}
\end{figure}

Once $\alpha$ decreases below $\alpha_{c1}\simeq0.2991$, multiple zero modes appear. The top panels of Fig.~\ref{fig:multi_zm} show the case of two zero modes, for $\alpha=0.28$. Both zero modes lie close to the center of the SBS spiral. The ring branch from the Newtonian limit ends at the first zero mode, and a short chain branch bifurcates on the other side of this zero mode, as in the case of a single zero mode. However, this chain branch now ends at the second zero mode, and a ring branch leaves the other side of the second zero mode toward the center of the SBS spiral. In other words, chains and rings alternate from one zero mode to the next. The relations between these solutions are seen most clearly from the quadrupole moment in the right panel. The chains between the two zero modes are only slightly deformed from spherical symmetry, so that their masses almost coincide with those of the SBSs in the left panel.

The case $\alpha=0.297$ extends this picture by one more zero mode. The ring branch from the second zero mode now ends at the third one, and a chain branch leaves the third zero mode toward the center of the spiral. In both cases, the chain branch from the Newtonian limit is detached. Thus, along an SBS family, each zero mode always has a chain branch on one side and a ring branch on the other, and their order alternates from one zero mode to the next.

Moreover, for all the values of $\alpha$ we have considered, the ring branch from the Newtonian limit ends at the first zero mode, whereas the chain branch from the Newtonian limit does so only for $\alpha\gtrsim0.38$. We suspect that this is because chains, unlike rings, cannot exist without gravity, so that the chain branch from the Newtonian limit is the first to detach as gravity becomes weaker compared with the self-interactions.

\begin{figure}[t]
  \centering
  \includegraphics[width=0.49\textwidth]{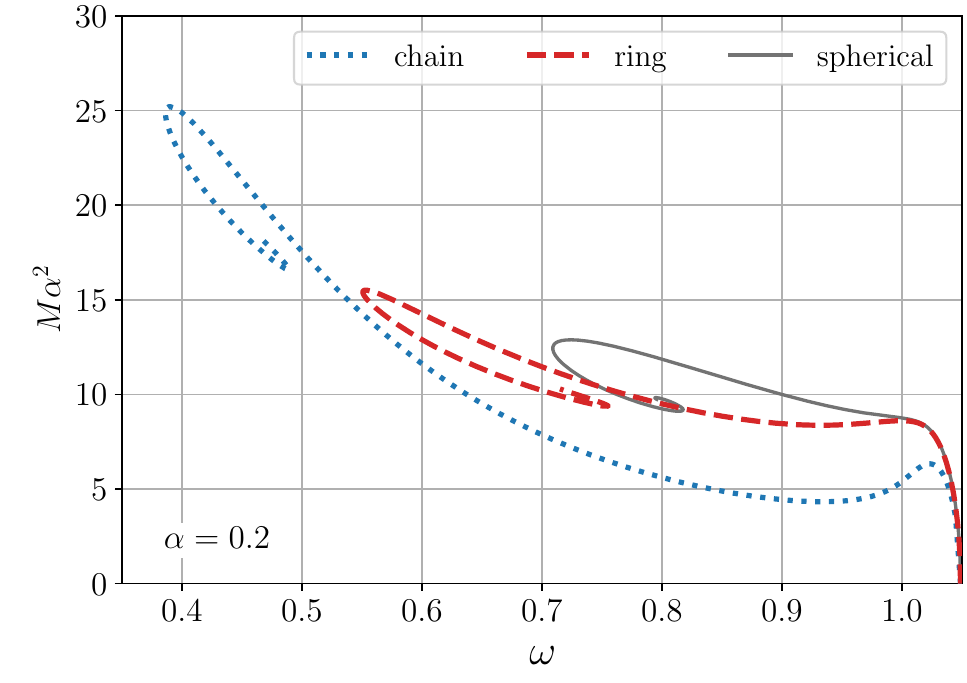}\hfill
  \includegraphics[width=0.49\textwidth]{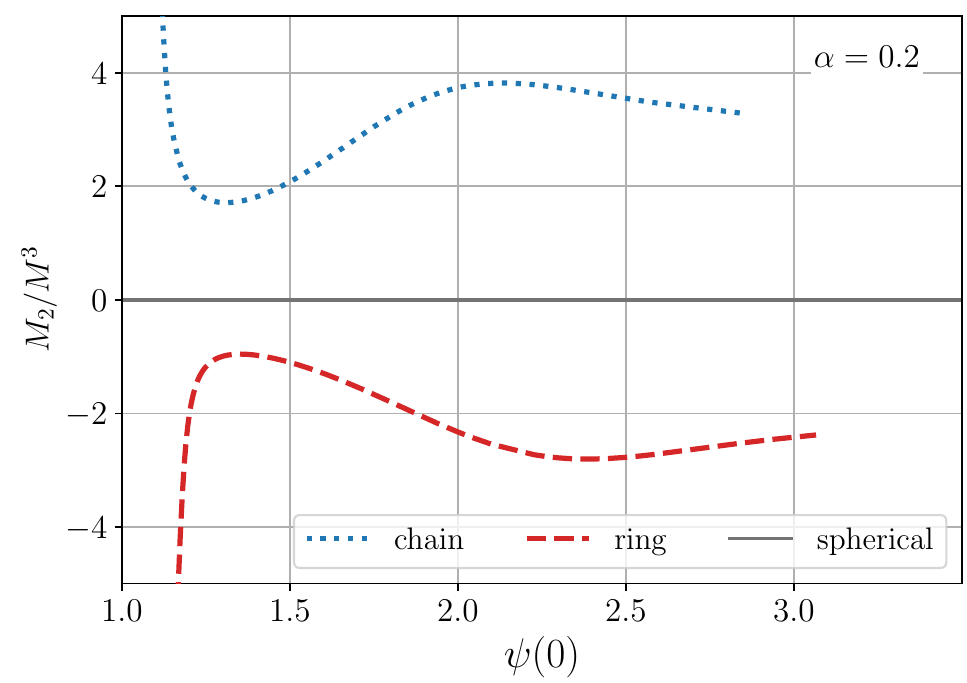}\\[4pt]
  \begin{minipage}[c]{0.49\textwidth}
    \includegraphics[width=\linewidth]{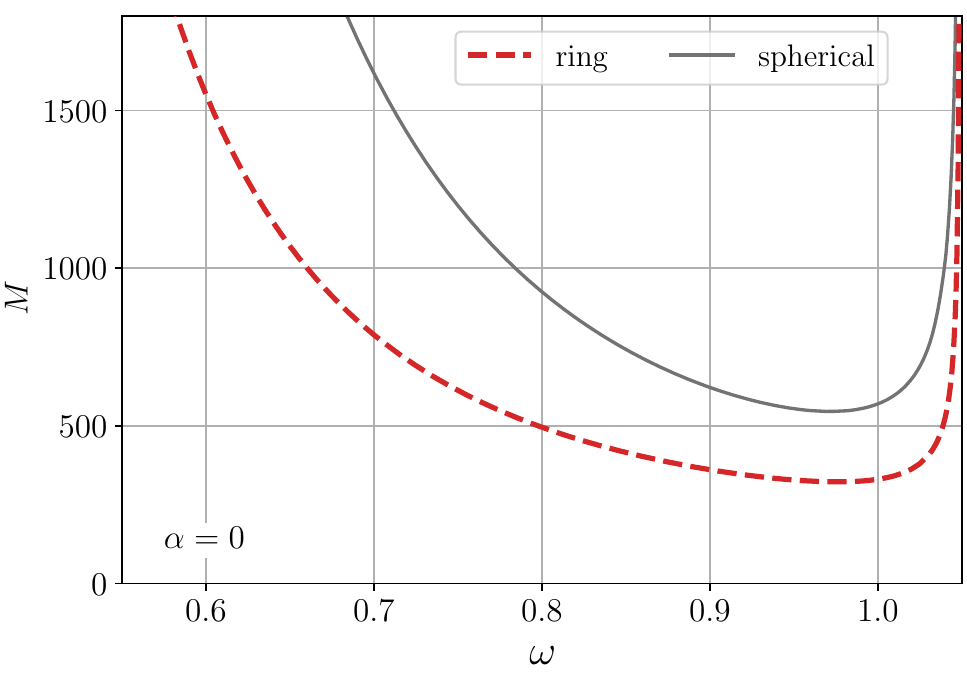}
  \end{minipage}\hfill
  \begin{minipage}[c]{0.45\textwidth}
    \centering
    \includegraphics[width=0.82\linewidth]{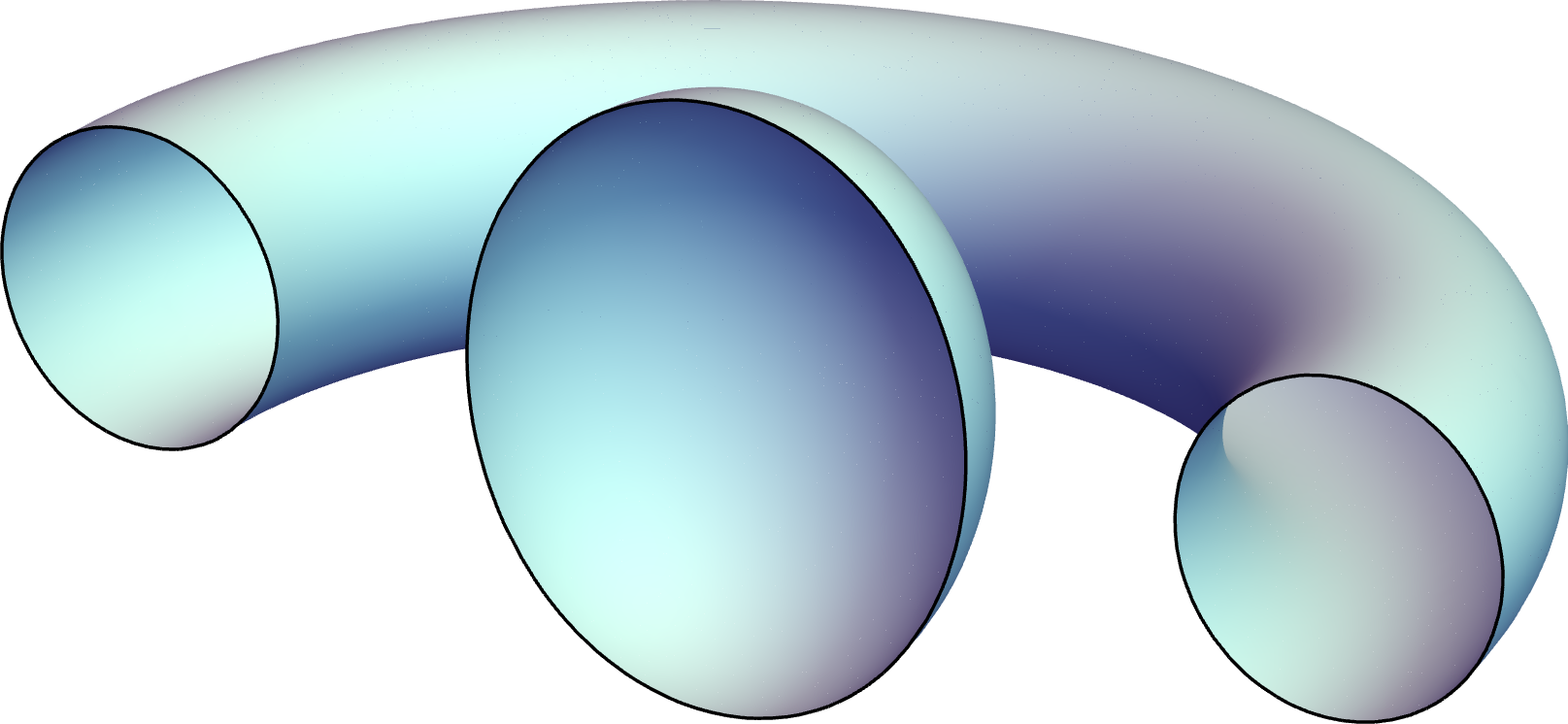}
  \end{minipage}
  \caption{Same as Fig.~\ref{fig:single_zm}, but for $\alpha=0.2$ (top), where the SBS family has no zero mode, and for $\alpha=0$ (bottom left), where no chains are found and the mass $M$ itself is shown. The bottom right panel shows the surface of constant energy density $\rho=0.3$ of the ring with $\alpha=0$ and $\omega=0.92$.}
  \label{fig:no_zm}
\end{figure}

For $\alpha<\alpha_{c2}$, the zero modes disappear completely. As shown in the top panels of Fig.~\ref{fig:no_zm}, for $\alpha=0.2$ only one chain branch and one ring branch remain. Both extend to the Newtonian limit, and both start to form a spiral, similar to that of the SBSs. The top right panel also shows that the bifurcations from the SBSs into ABSs have disappeared. 

Interestingly, in flat spacetime, where there is no gravitational binding, we find no chains, whereas the rings survive, as shown in the bottom left panel. For $\alpha=0$, the masses of the spherical Q-balls with one node and of the rings diverge both as $\omega\to\mu_0$ and at small $\omega$. At a given frequency, the mass of a ring is only about half that of the spherical Q-ball. The bottom right panel shows a surface of constant energy density of the ring with $\omega=0.92$, which illustrates the matter distribution of this type of soliton. To check the accuracy of the rings in flat spacetime, we use the virial identity $\int|\nabla\psi|^2\,d^3x=3\int(\omega^2\psi^2-U)\,d^3x$, which follows from Derrick's scaling argument~\cite{Derrick:1964ww,Herdeiro:2022ids}. The rings satisfy it with a relative error of about $10^{-4}$.

In this model, a static axisymmetric Q-ball with the same parity as the rings was constructed in Ref.~\cite{Brihaye:2007tn}, as an angular excitation of the fundamental Q-ball related to the spherical harmonic $Y_2^0$. According to Ref.~\cite{Brihaye:2007tn}, its energy density has two local maxima on the positive symmetry axis, unlike that of the rings.

\section{Conclusions}\label{sec:conclusions}
In this work, we have studied how self-interactions affect the $\ell=2$ zero modes of spherical boson stars (SBSs) with one radial node, at which the SBSs bifurcate into chains and rings. We considered massive BSs, with a repulsive quartic self-interaction, and solitonic BSs, with a Q-ball type potential containing quartic and sextic terms. In both models, we followed the zero modes as the couplings vary, and constructed the chains and rings that bifurcate from them.

For massive BSs, the zero mode is not removed by the quartic self-interaction, at least up to $\lambda\approx3000$. As $\lambda$ increases, the frequency of the SBS at the zero mode moves toward the Newtonian limit, while that of the maximum-mass solution decreases. The two frequencies cross at $\lambda\approx15$, and for larger $\lambda$, the zero mode lies on the candidate stable branch, between the maximum-mass solution and the Newtonian limit. This branch therefore cannot be entirely stable against $\ell=2$ perturbations, although strong enough self-interactions quench the instability of the excited SBSs in spherically symmetric evolutions~\cite{Sanchis-Gual:2021phr,Brito:2023fwr}. This is consistent with the non-spherical instability found in 3+1 evolutions~\cite{Brito:2025rld}.

For solitonic BSs, the bifurcations depend strongly on the gravitational coupling $\alpha$. As $\alpha$ decreases, the chain branch from the Newtonian limit detaches from the zero mode at $\alpha\approx0.38$, although a chain branch still bifurcates there together with the ring branch. At smaller $\alpha$, the zero modes trace a spiral in the $(1/\alpha^2,\omega)$ plane, so that for $\alpha_{c2}<\alpha<\alpha_{c1}$, with $\alpha_{c1}\simeq0.2991$ and $\alpha_{c2}\simeq0.2756$, a single SBS family has two or more zero modes. Along such a family, chain and ring branches alternate from one zero mode to the next. For $\alpha<\alpha_{c2}$, the zero modes disappear, and the chain and ring branches are no longer connected to the SBSs. In flat spacetime, we still find rings, as static ring-like Q-balls, but no chains. This suggests that chains, unlike rings, need gravity to exist.

A natural extension is to consider time-dependent perturbations instead of the static ones used here, which would show on which side of a zero mode the unstable mode lies and how fast it grows. Moreover, a rigorous proof that chains, unlike rings, cannot exist without gravity is still missing, since the argument of Ref.~\cite{Cunha:2022tvk} only excludes static parity-odd configurations in flat spacetime. In addition, SBSs with more nodes have further zero modes, including ones under $\ell=4$ perturbations~\cite{Liang:2025myf}, and it would be interesting to see how self-interactions affect them. Finally, spherical Proca stars~\cite{Brito:2015pxa} also have an $\ell=2$ zero mode~\cite{Herdeiro:2023wqf,Liang:2025myf}. Since self-interacting Proca fields can suffer from a breakdown of hyperbolicity~\cite{Clough:2022ygm,Coates:2022qia}, the effect of self-interactions on this zero mode could be studied through the $\ell=2$ perturbations of Proca-Higgs stars~\cite{Herdeiro:2023lze,Brito:2024biy}. These stars reduce to mini-Proca stars in some limits, so the zero mode should persist for them, at least close to these limits.

\section*{Acknowledgments}
I thank C. Herdeiro and E. Radu for their advice and support. I also gratefully acknowledge the computational resources of the Argus cluster at the University of Aveiro. This work is supported by the Centre for Research and Development in Mathematics and Applications (CIDMA, \url{https://ror.org/05pm2mw36}) under the Portuguese Foundation for Science and Technology (FCT -- Fundaç\~ao para a Ci\^encia e a Tecnologia, \url{https://ror.org/00snfqn58}), Grants UID/04106/2025 (\url{https://doi.org/10.54499/UID/04106/2025}) and UID/PRR/04106/2025 (\url{https://doi.org/10.54499/UID/PRR/04106/2025}), as well as by the Horizon Europe staff exchange (SE) programme HORIZON-MSCA-2021-SE-01, Grant No.\ NewFunFiCO-101086251, and by the project 2022.04560.PTDC (\url{https://doi.org/10.54499/2022.04560.PTDC}). The author is supported by a China Scholarship Council (CSC) fellowship. 

\bibliographystyle{unsrt}
\bibliography{refs}

\end{document}